\documentclass[aps,prc,preprint,superscriptaddress]{revtex4}
\usepackage[utf8]{inputenc}
\usepackage[T1]{fontenc}
\usepackage{lmodern,textcomp}
\usepackage{setspace}
\usepackage{microtype}
\usepackage{epigraph}
\usepackage{natbib}
\usepackage{graphicx}
\usepackage{array}
\usepackage{amsmath} 
\usepackage{multirow}

\begin{document}

\title{Measurement of deeply virtual Compton scattering off Helium-4 with CLAS at Jefferson Lab}

\newcommand*{\ANL}{Argonne National Laboratory, Argonne, Illinois 60439}
\newcommand*{\ANLindex}{1}
\affiliation{\ANL}
\newcommand*{\CANISIUS}{Canisius College, Buffalo, NY}
\newcommand*{\CANISIUSindex}{2}
\affiliation{\CANISIUS}
\newcommand*{\CMU}{Carnegie Mellon University, Pittsburgh, Pennsylvania 15213}
\newcommand*{\CMUindex}{3}
\affiliation{\CMU}
\newcommand*{\SACLAY}{IRFU, CEA, Universit\'{e} Paris-Saclay, F-91191 Gif-sur-Yvette, France}
\newcommand*{\SACLAYindex}{4}
\affiliation{\SACLAY}
\newcommand*{\CNU}{Christopher Newport University, Newport News, Virginia 23606}
\newcommand*{\CNUindex}{5}
\affiliation{\CNU}
\newcommand*{\UCONN}{University of Connecticut, Storrs, Connecticut 06269}
\newcommand*{\UCONNindex}{6}
\affiliation{\UCONN}
\newcommand*{\DUKE}{Duke University, Durham, North Carolina 27708-0305}
\newcommand*{\DUKEindex}{7}
\affiliation{\DUKE}
\newcommand*{\DUQUESNE}{Duquesne University, 600 Forbes Avenue, Pittsburgh, PA 15282 }
\newcommand*{\DUQUESNEindex}{8}
\affiliation{\DUQUESNE}
\newcommand*{\FU}{Fairfield University, Fairfield CT 06824}
\newcommand*{\FUindex}{9}
\affiliation{\FU}
\newcommand*{\FERRARAU}{Universita' di Ferrara , 44121 Ferrara, Italy}
\newcommand*{\FERRARAUindex}{10}
\affiliation{\FERRARAU}
\newcommand*{\FIU}{Florida International University, Miami, Florida 33199}
\newcommand*{\FIUindex}{11}
\affiliation{\FIU}
\newcommand*{\FSU}{Florida State University, Tallahassee, Florida 32306}
\newcommand*{\FSUindex}{12}
\affiliation{\FSU}
\newcommand*{\GWUI}{The George Washington University, Washington, DC 20052}
\newcommand*{\GWUIindex}{13}
\affiliation{\GWUI}
\newcommand*{\ISU}{Idaho State University, Pocatello, Idaho 83209}
\newcommand*{\ISUindex}{14}
\affiliation{\ISU}
\newcommand*{\INFNFE}{INFN, Sezione di Ferrara, 44100 Ferrara, Italy}
\newcommand*{\INFNFEindex}{15}
\affiliation{\INFNFE}
\newcommand*{\INFNFR}{INFN, Laboratori Nazionali di Frascati, 00044 Frascati, Italy}
\newcommand*{\INFNFRindex}{16}
\affiliation{\INFNFR}
\newcommand*{\INFNGE}{INFN, Sezione di Genova, 16146 Genova, Italy}
\newcommand*{\INFNGEindex}{17}
\affiliation{\INFNGE}
\newcommand*{\INFNRO}{INFN, Sezione di Roma Tor Vergata, 00133 Rome, Italy}
\newcommand*{\INFNROindex}{18}
\affiliation{\INFNRO}
\newcommand*{\INFNTUR}{INFN, Sezione di Torino, 10125 Torino, Italy}
\newcommand*{\INFNTURindex}{19}
\affiliation{\INFNTUR}
\newcommand*{\INFNPAV}{INFN, Sezione di Pavia, 27100 Pavia, Italy}
\newcommand*{\INFNPAVindex}{20}
\affiliation{\INFNPAV}
\newcommand*{\ORSAY}{Universit\'{e} Paris-Saclay, CNRS/IN2P3, IJCLab, 91405 Orsay, France}
\newcommand*{\ORSAYindex}{21}
\affiliation{\ORSAY}
\newcommand*{\Juelich}{Institute fur Kernphysik (Juelich), Juelich, Germany}
\newcommand*{\Juelichindex}{22}
\affiliation{\Juelich}
\newcommand*{\KNU}{Kyungpook National University, Daegu 41566, Republic of Korea}
\newcommand*{\KNUindex}{23}
\affiliation{\KNU}
\newcommand*{\LAMAR}{Lamar University, 4400 MLK Blvd, PO Box 10046, Beaumont, Texas 77710}
\newcommand*{\LAMARindex}{24}
\affiliation{\LAMAR}
\newcommand*{\LPSC}{LPSC, Universit\'e Grenoble-Alpes, CNRS/IN2P3, 38026 
Grenoble, France}
\newcommand*{\LPSCindex}{25}
\affiliation{\LPSC}
\newcommand*{\MIT}{Massachusetts Institute of Technology, Cambridge, Massachusetts  02139-4307}
\newcommand*{\MITindex}{26}
\affiliation{\MIT}
\newcommand*{\MISS}{Mississippi State University, Mississippi State, MS 39762-5167}
\newcommand*{\MISSindex}{27}
\affiliation{\MISS}
\newcommand*{\ITEP}{National Research Centre Kurchatov Institute - ITEP, Moscow, 117259, Russia}
\newcommand*{\ITEPindex}{28}
\affiliation{\ITEP}
\newcommand*{\UNH}{University of New Hampshire, Durham, New Hampshire 03824-3568}
\newcommand*{\UNHindex}{29}
\affiliation{\UNH}
\newcommand*{\NMSU}{New Mexico State University, PO Box 30001, Las Cruces, NM 88003, USA}
\newcommand*{\NMSUindex}{30}
\affiliation{\NMSU}
\newcommand*{\NSU}{Norfolk State University, Norfolk, Virginia 23504}
\newcommand*{\NSUindex}{31}
\affiliation{\NSU}
\newcommand*{\OHIOU}{Ohio University, Athens, Ohio  45701}
\newcommand*{\OHIOUindex}{32}
\affiliation{\OHIOU}
\newcommand*{\ODU}{Old Dominion University, Norfolk, Virginia 23529}
\newcommand*{\ODUindex}{33}
\affiliation{\ODU}
\newcommand*{\JLUGiessen}{II Physikalisches Institut der Universitaet Giessen, 35392 Giessen, Germany}
\newcommand*{\JLUGiessenindex}{34}
\affiliation{\JLUGiessen}
\newcommand*{\URICH}{University of Richmond, Richmond, Virginia 23173}
\newcommand*{\URICHindex}{35}
\affiliation{\URICH}
\newcommand*{\ROMAII}{Universita' di Roma Tor Vergata, 00133 Rome Italy}
\newcommand*{\ROMAIIindex}{36}
\affiliation{\ROMAII}
\newcommand*{\MSU}{Skobeltsyn Institute of Nuclear Physics, Lomonosov Moscow State University, 119234 Moscow, Russia}
\newcommand*{\MSUindex}{37}
\affiliation{\MSU}
\newcommand*{\SCAROLINA}{University of South Carolina, Columbia, South Carolina 29208}
\newcommand*{\SCAROLINAindex}{38}
\affiliation{\SCAROLINA}
\newcommand*{\TEMPLE}{Temple University,  Philadelphia, PA 19122 }
\newcommand*{\TEMPLEindex}{39}
\affiliation{\TEMPLE}
\newcommand*{\JLAB}{Thomas Jefferson National Accelerator Facility, Newport News, Virginia 23606}
\newcommand*{\JLABindex}{40}
\affiliation{\JLAB}
\newcommand*{\UTFSM}{Universidad T\'{e}cnica Federico Santa Mar\'{i}a, Casilla 110-V Valpara\'{i}so, Chile}
\newcommand*{\UTFSMindex}{41}
\affiliation{\UTFSM}
\newcommand*{\INSUBRIA}{Universit\`{a} degli Studi dell'Insubria, 22100 Como, Italy}
\newcommand*{\INSUBRIAindex}{42}
\affiliation{\INSUBRIA}
\newcommand*{\BRESCIA}{Universit\`{a} degli Studi di Brescia, 25123 Brescia, Italy}
\newcommand*{\BRESCIAindex}{43}
\affiliation{\BRESCIA}
\newcommand*{\GLASGOW}{University of Glasgow, Glasgow G12 8QQ, United Kingdom}
\newcommand*{\GLASGOWindex}{44}
\affiliation{\GLASGOW}
\newcommand*{\YORK}{University of York, York YO10 5DD, United Kingdom}
\newcommand*{\YORKindex}{45}
\affiliation{\YORK}
\newcommand*{\VIRGINIA}{University of Virginia, Charlottesville, Virginia 22901}
\newcommand*{\VIRGINIAindex}{46}
\affiliation{\VIRGINIA}
\newcommand*{\VT}{Virginia Tech, Blacksburg, Virginia   24061-0435}
\newcommand*{\VTindex}{47}
\affiliation{\VT}
\newcommand*{\WM}{College of William and Mary, Williamsburg, Virginia 23187-8795}
\newcommand*{\WMindex}{48}
\affiliation{\WM}
\newcommand*{\YEREVAN}{Yerevan Physics Institute, 375036 Yerevan, Armenia}
\newcommand*{\YEREVANindex}{49}
\affiliation{\YEREVAN}
 
 %%%%%%%%%%%%%%% END OF Latex Macros for institute addresses  
%%%%%%%%%%%%%%%%%%%%%%%%%% 

\author {R.~Dupr\'{e}} 
\email[corresponding author: ]{raphael.dupre@ijclab.in2p3.fr}
\affiliation{\ANL}
\affiliation{\ORSAY}
\author {M.~Hattawy}
\affiliation{\ANL}
\affiliation{\ORSAY}
\affiliation{\ODU}
\author {N.A.~Baltzell} 
\affiliation{\ANL}
\affiliation{\JLAB}
\author {S.~B\"{u}ltmann} 
\affiliation{\ODU}
\author{R.~De~Vita} 
\affiliation{\INFNGE}
\author {A.~El~Alaoui} 
\affiliation{\ANL}
\affiliation{\UTFSM}
\author {L.~El~Fassi} 
\affiliation{\ANL}
\affiliation{\MISS}
\author{H.~Egiyan}
\affiliation{\JLAB}
\author{F.X.~Girod} 
\affiliation{\JLAB}
\author {M.~Guidal} 
\affiliation{\ORSAY}
% ...

\collaboration{The CLAS Collaboration}
\noaffiliation

\date{\today}

\begin{abstract}
We report on the measurement of the beam spin asymmetry in the deeply virtual Compton 
scattering off $^4$He using the CEBAF Large Acceptance Spectrometer (CLAS) at
Jefferson Lab using a 6 GeV longitudinally polarized electron beam incident on a 
pressurized $^4$He gaseous target. We detail the method used to ensure the exclusivity 
of the measured reactions, in particular the upgrade of CLAS with a radial time 
projection chamber to detect the 
low-energy recoiling $^4$He nuclei and an inner calorimeter to extend the 
photon detection acceptance at forward angles. Our results confirm the 
theoretically predicted enhancement of the coherent 
($e^4$He~$\to~e'$$^4$He$'\gamma'$) beam spin asymmetries compared to those 
observed on the free proton, while the incoherent 
($e^4$He~$\to~e'$p$'\gamma'$X$'$) asymmetries exhibit a 30$\%$ suppression.  
From the coherent data, we were able to extract, in a model-independent way, 
the real and imaginary parts of the only $^4$He Compton form factor, $\cal 
H_A$, leading the way toward 3D imaging of the partonic structure of nuclei.
\end{abstract}

%\pacs{}

\maketitle

\section{Introduction}

In the past few decades, the study of the proton structure has made significant progress
thanks to the theoretical and experimental developments of three dimensional 
structure functions \cite{Anselmino:2015uka}. These studies, which have focused on generalized
parton distributions (GPDs) and transverse momentum dependent parton distribution functions (TMDs)
can be generalized to the nucleus and offer a unique opportunity to revisit the quark structure 
of the nucleus with an original perspective~\cite{Dupre:2015jha}. This new approach is 
particularly needed as the quark structure of the nucleus remains today the subject of numerous
controversies. Indeed, while much progress has been made in measuring the nuclear
parton distribution functions, their shape can be explained with very different model 
assumptions~\cite{Norton:2003cb,Malace:2014uea,Hen:2016kwk}.

In nuclei, the GPDs can be probed conveniently through the measurement of the
spin asymmetries generated by the deeply virtual Compton scattering (DVCS) 
process~\cite{Diehl:2003ny,Belitsky:2005qn,Boffi:2007yc,Guidal:2013rya}. The measurement of the exclusive
production of a photon limits the possibilities of final state interactions (FSIs) in the nuclear 
medium and offers a unique opportunity to make a measurement free of them. 
Moreover, with a spin-0 nuclear target, the extraction of the GPD from the DVCS data 
is significantly simplified since a single GPD is involved in the process at leading order. However, the 
measurement of the nuclear DVCS is challenging experimentally and the first attempts by 
the HERMES Collaboration \cite{Airapetian:2009cga} to unravel an $A$ dependent nuclear
effect have been unsuccessful. We present here
in detail the more recent measurements by the CLAS Collaboration, which has been already partially presented
in two short letters \cite{Hattawy:2017woc,Hattawy:2018liu}. We extend in this article 
the description of the CLAS nuclear DVCS experiment, detail the methods used for the
data analysis and produce the complete experimental results for each channel measured.

\section{Theoretical Framework}

\subsection{The GPD Formalism}

\begin{figure}[tbp!]
\center
\includegraphics[width=10.0cm]{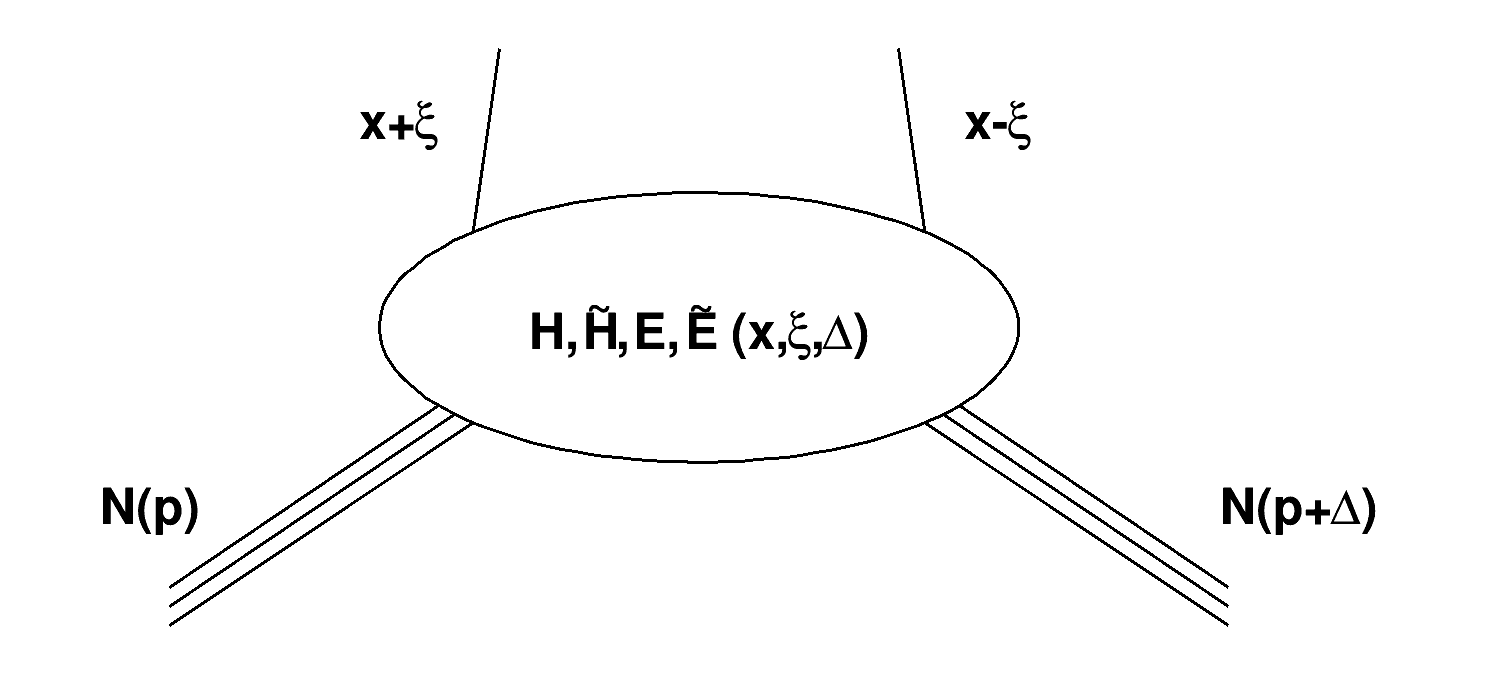}
\caption{General representation for the GPDs of a nucleon represented by the triple lines and noted $N$.
	Single lines can represent quarks or anti-quarks probed in the nucleon shown by the triple lines.}
\label{fig:GPD}
\end{figure}

The theory of GPDs has been already reviewed in detail in various 
publications~\cite{Diehl:2003ny,Belitsky:2005qn,Boffi:2007yc,Guidal:2013rya},
and we summarize here only the necessary elements to discuss the present 
experimental results. The GPDs are real structure functions $F^{q}(x,\xi,t)$, 
where $x+\xi$ and $x-\xi$ are the incoming and outgoing quark momenta respectively 
and $t=\Delta^2$ is the squared transferred four momentum to the target, as 
illustrated in Fig.~\ref{fig:GPD}. 

The different possible spin states lead to several independent GPDs for any 
given hadron. The proper accounting of the number of GPDs must be done with 
regard to the symmetries of the system. At leading order and leading twist, we 
find that there are $2(2J+1)^2$ GPDs for a particle of spin $J$. 
Therefore for a spin-0 hadron like the helium-4 nucleus, we will have two GPDs, and 
for a spin-1/2 hadron like the proton, eight GPDs. Half of these involve a 
parton helicity flip, they are called transversity GPDs and do not contribute to 
the DVCS process. 

DVCS is the main experimental probe of the GPDs. However, this process does not allow
for an extraction of the GPDs in the full phase space of the parameters. Instead, 
DVCS gives access to the GPDs integrated over $x$. 
To account for this and simplify the notation, we define the complex Compton form factors (CFF,
noted with curved $\mathcal{F}$ for a given GPD $F$) for each GPD as follows:
\begin{equation}
\Re e (\mathcal{F}(\xi,t)) = \sum _q e^2_q \mathcal{P} \int^1_{-1} dx F^{q}(x,\xi,t)
    \left [ {1 \over x - \xi} \mp {1 \over x + \xi} \right], 
\end{equation}
\begin{equation}
\Im m (\mathcal{F}(\xi,t)) = - \pi \sum _q e^2_q \left [ F^{q}(\xi,\xi,t) \mp F^{q}(-\xi,\xi,t) \right]. 
\end{equation} 
These are the quantities directly present in the DVCS cross sections. We note that they are summed over
the different quark flavors present in the hadron, as the electromagnetic probe does not differentiate
quark flavors.

Experimentally, another process is indistinguishable from DVCS, the Bethe-Heitler (BH) 
process in which 
the final state photon is emitted by the scattering lepton rather than the hadron. In this case, the 
photon-hadron interaction is the same as in elastic scattering and depends on the
target form factors rather than its GPDs. The DVCS and BH processes are experimentally 
indistinguishable as they
have identical final states, such that they interfere in the squared amplitude of the
exclusive photo-production process:
\begin{equation}
|T|^2 = |T_{DVCS}|^2 + |T_{BH}|^2 + T_{DVCS}^* T_{BH} + T_{DVCS}T_{BH}^*. 
\end{equation}
The interference terms significantly increase the cross section in specific parts of the phase space
and lead to significant beam spin asymmetries (BSAs), which are the focus of the measurements presented 
here. 

\begin{figure}[tbp!]
\center
\includegraphics[width=9cm]{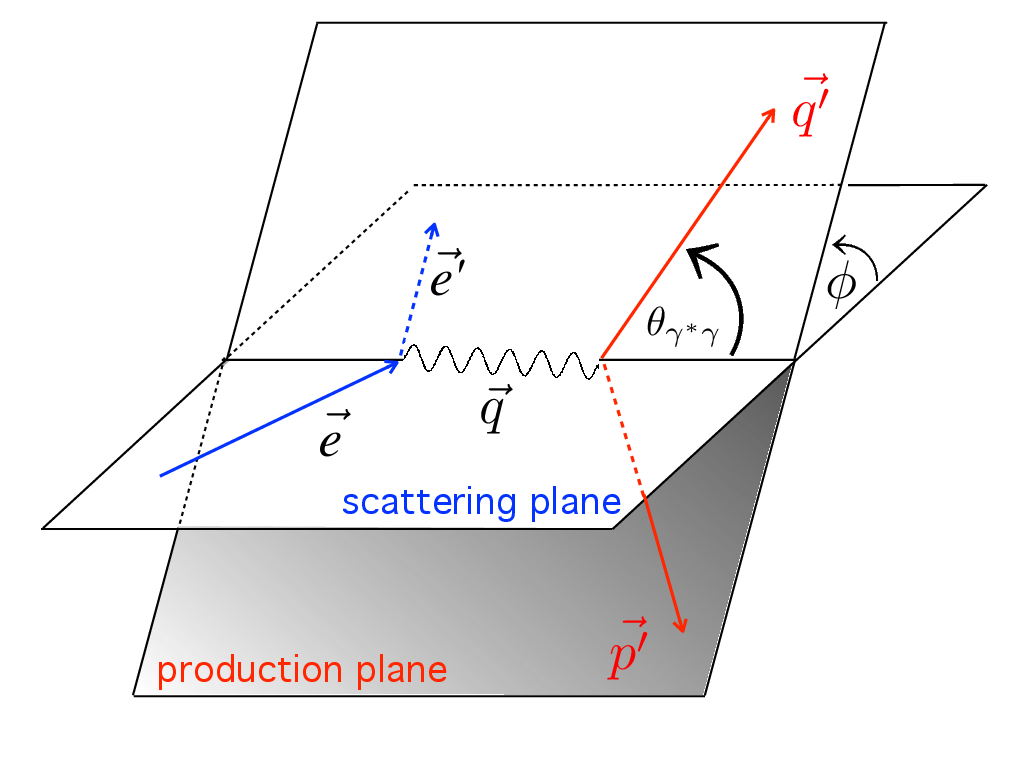}
	\caption{Illustration of the scattering (or leptonic) and production (or hadronic) planes
	in the DVCS process.} 
\label{fig:PhiAngle}
\end{figure}

Finally, we need to define the kinematics. We use the 
conventions from Fig. \ref{fig:PhiAngle} for angles and the experimental kinematic variables
used here are defined as: 
$-t = -(p_p-p_p^\prime)^2 = \Delta^2$ and 
$x_B = {Q^2 \over 2 M_N \nu}\sim {2\xi \over 1+\xi}$, with $M_N$ the nucleon mass and $\nu$ the
energy transfer to the target, $\nu = {E -E^\prime} $. 

\subsection{Coherent Nuclear DVCS}

\begin{figure}[tbp!]
\center
\includegraphics[width=9cm]{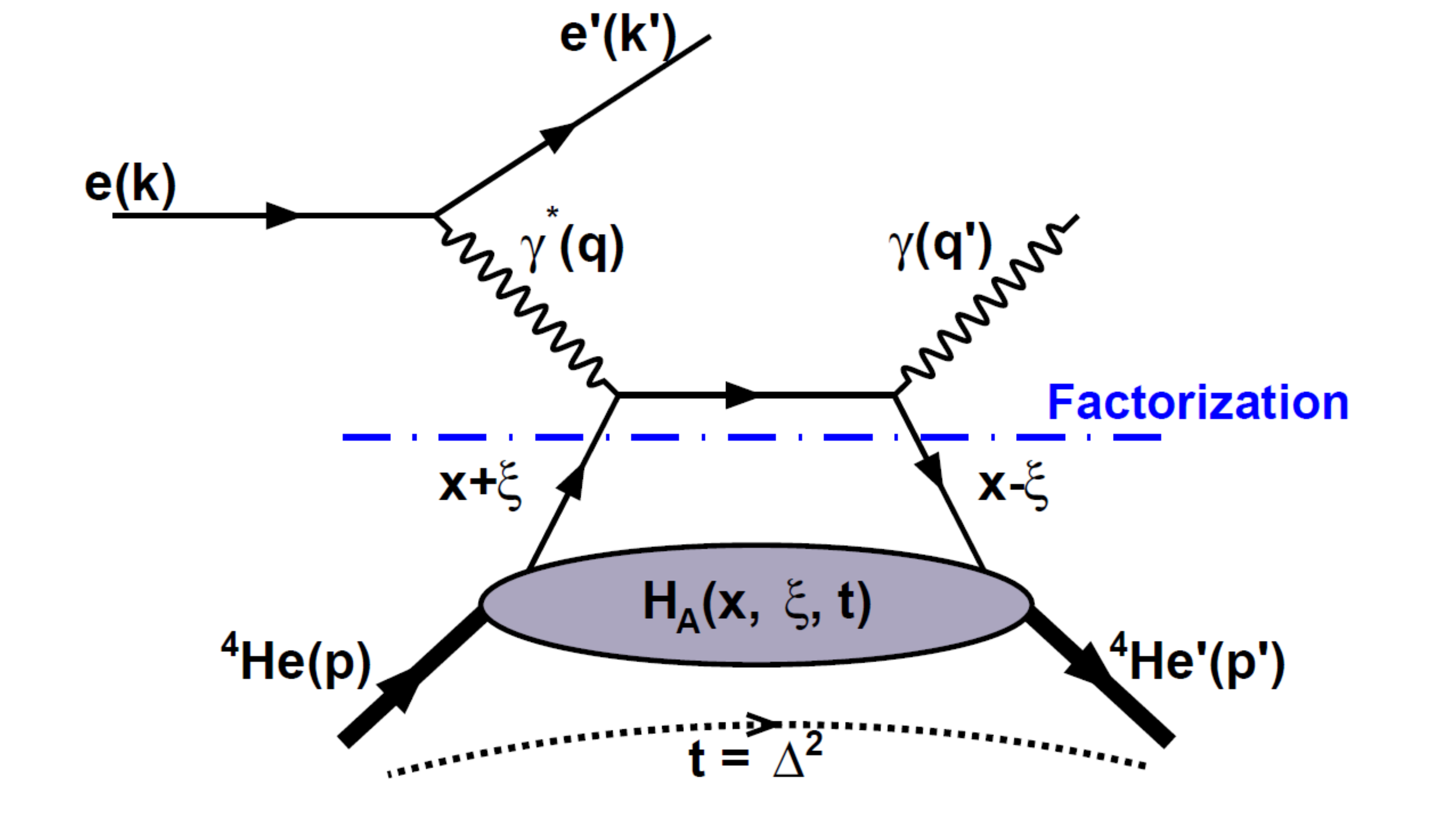}
\caption{Diagram representing the coherent nuclear DVCS, where we 
indicate the limit between the hard and the soft components with the dot-dashed factorization line.}
\label{fig:CohDiag}
\end{figure}

The first reaction measured in the experiment is the coherent electro-production of a photon on helium
$e+^4\!\!He \rightarrow e^\prime+\gamma+^4\!\!He^\prime$ at large 4-momentum transfer squared ($Q^2$). The 
leading order diagram of the nuclear coherent DVCS is represented in Fig. \ref{fig:CohDiag}.
In the present experiment, we focused on the measurement of the BSA 
noted $A_{LU}$ with 
$L$ for the longitudinally polarized electron beam and $U$ the unpolarized target, which is defined as:  
\begin{equation}
A_{LU} = \frac{d^{5}\sigma^{+} - d^{5}\sigma^{-} }
              {d^{5}\sigma^{+} + d^{5}\sigma^{-}},
  \label{eq:BSA}
\end{equation}
where $d^{5}\sigma^{+}$($d^{5}\sigma^{-}$) is the differential cross section for a positive 
(negative) beam helicity. At leading order and leading twist, the BSA can be expressed as \cite{Kirchner:2003wt}:        
\begin{eqnarray}
\label{eq:coh_BSA}
A_{LU}& =& \frac{x_A(1+\epsilon^2)^2}{y} \, s_1^{INT} \sin(\phi) \, 
\bigg/ \, \bigg[ \, \sum_{n=0}^{n=2}c_n^{BH}\cos{(n\phi)} +  \\
& & \frac{x_A^2 t {(1+\epsilon^2)}^2}{Q^2} P_1(\phi) P_2(\phi) \, c_0^{DVCS} + 
\frac{x_A (1+\epsilon^2)^2}{y} \sum_{n=0}^{n=1} c_n^{INT} \cos{(n\phi)} \bigg],  \nonumber 
\end{eqnarray}
where $\mathcal{P}_1(\phi)$ and $\mathcal {P}_2(\phi)$ are the BH propagators, 
and $x_{A} = \frac{M_{p}\cdot x}{M_{^4\!He}}$. The factors: $c_{0,1,2}^{BH}$, 
$c_0^{DVCS}$, $c_{0,1}^{INT}$ and 
$s_1^{INT}$ are the Fourier coefficients of the BH, the DVCS and the 
interference amplitudes for a spin-zero target, respectively. The explicit 
expressions of these coefficients, which have been derived based on the work of 
Kirchner and Müller \cite{Kirchner:2003wt}, can be found in Appendix \ref{sec:eq}.

This formula can be expressed in a simplified manner for a spin-0 target as \cite{Belitsky:2008bz}:
\begin{equation}
A_{LU}(\phi) = \frac{\alpha_{0}(\phi) \, \Im m(\mathcal{H}_{A})}
{\alpha_{1}(\phi) + \alpha_{2}(\phi) \, \Re e(\mathcal{H}_{A}) + \alpha_{3}(\phi) \, 
\big( 
\Re e(\mathcal{H}_{A})^{2} + \Im m(\mathcal{H}_{A})^{2} \big)},
\label{eq:A_LU-coh}
\end{equation}
where $\Im m(\mathcal{H}_{A})$ and $\Re e(\mathcal{H}_{A})$ are the imaginary 
and real parts,respectively, of the CFF $\mathcal{H}_{A}$ associated with 
the GPD $H_A$ of the spin-0 nucleus. The 
$\alpha_{i}$ factors are $\phi$-dependent kinematical terms that depend on the 
nuclear form factor $F_A$ and the independent variables $Q^2$, $x$ and $t$.  
These factors have the following simplified expressions:
\begin{eqnarray}
	\label{eq:alpha1}
   \alpha_0 (\phi) & = &\frac{x_{A}(1+\epsilon^2)^2}{y} S_{++}(1) \sin(\phi)\\
    \alpha_1 (\phi) & = & c_0^{BH}+c_1^{BH} \cos({\phi})+c_2^{BH} \cos(2\phi)\\ 
   \alpha_2 (\phi) & = & \frac{x_{A}(1+\epsilon^2)^2}{y}  \left( C_{++}(0) +  
C_{++}(1) \cos(\phi) \right)\\
\alpha_3 (\phi) &=& \frac{x^{2}_{A}t(1+\epsilon^2)^2}{y} {\mathcal P}_1(\phi) 
{\mathcal P}_2(\phi) \cdot 2 \frac{2-2y+y^2 + \frac{\epsilon^2}{2}y^2}{1 + 
\epsilon^2},
	\label{eq:alpha4}
\end{eqnarray}
where $S_{++}(1)$, $C_{++}(0)$, and $C_{++}(1)$ are the Fourier harmonics found in the 
leptonic tensor \cite{Belitsky:2008bz}. Their explicit expression are provided in 
Appendix \ref{sec:eq}. 

\begin{figure}[tbp!]
\center
\includegraphics[width=8cm]{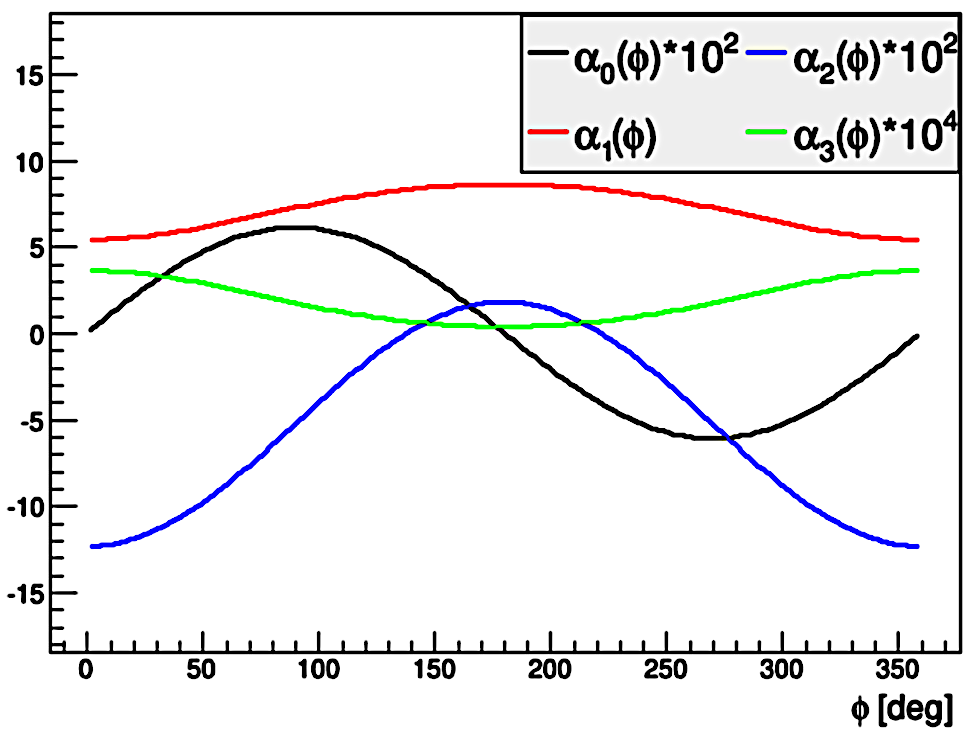}
	\caption{Coefficients presented in Eqs. \ref{eq:alpha1} to \ref{eq:alpha4}.
	Note the prescaling factors used for $\alpha_0$, $\alpha_2$ and $\alpha_3$.}
\label{fig:alphas}
\end{figure}

Eq. \ref{eq:A_LU-coh} is particularly convenient to perform an extraction of 
$\Im m(\mathcal{H}_{A})$ and $\Re e(\mathcal{H}_{A})$ through a fit of the BSA as 
a function of $\phi$. As can be seen in Fig. \ref{fig:alphas}, the form of each 
$\alpha$ coefficient has a characteristic $\phi$ dependence, such that a fit can
easily separate their respective contributions. The only caveat
is the large difference of magnitude between the $\alpha$ factors, which can lead to 
rather different error propagation for the two parts of the CFF.

An important issue with the use of this theoretical framework is the large mass of the 
helium nucleus. Recent work indicates that the effect of this correction is moderate 
\cite{Braun:2012bg}, however the applicability to such a large mass remains to be
fully explored from the theoretical point of view.

\subsection{Incoherent Nuclear DVCS}

\begin{figure}[tbp!]
\center
\includegraphics[width=10cm]{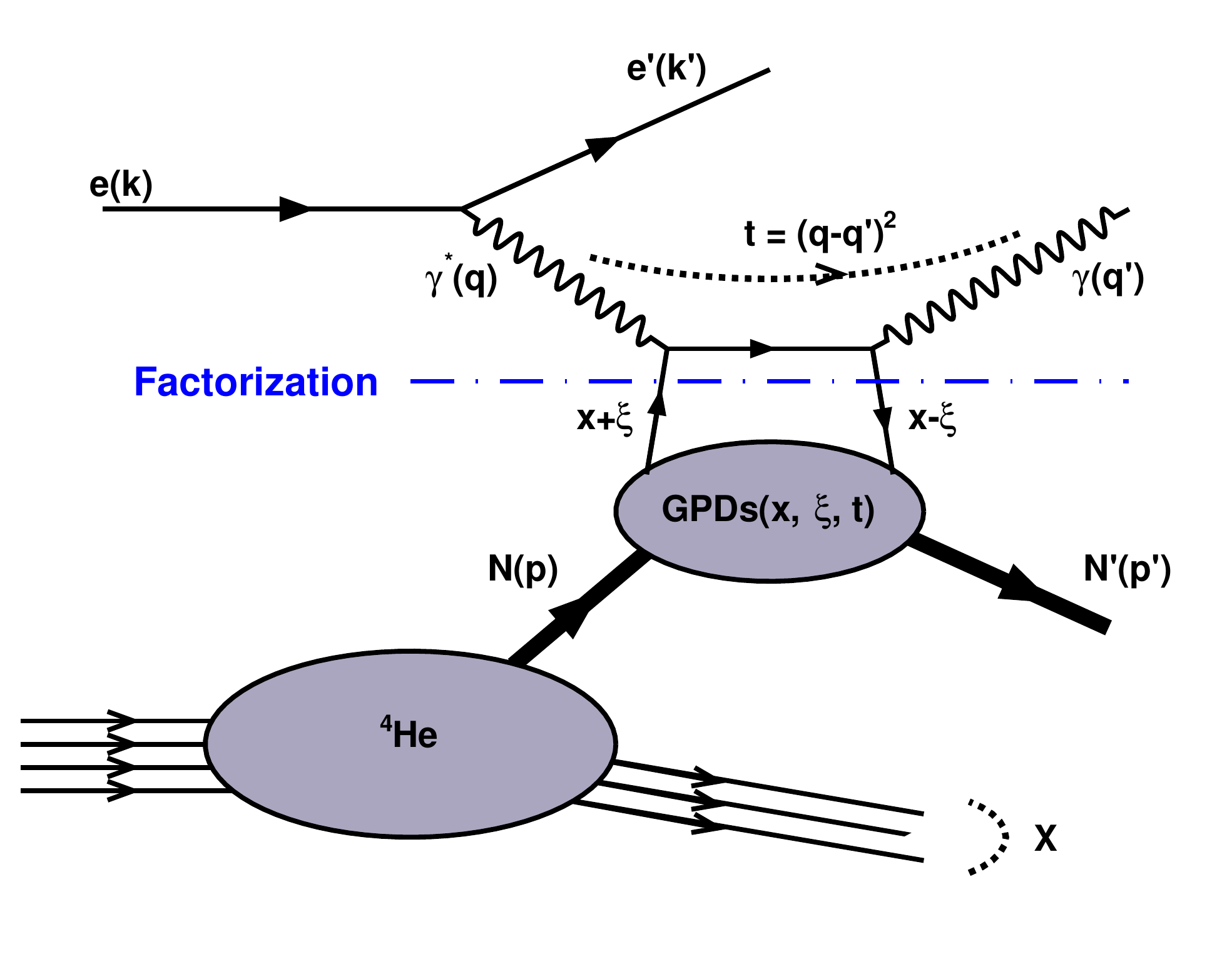}
\caption{Diagram representing the incoherent nuclear DVCS.}
\label{fig:InCohDiag}
\end{figure}

The incoherent nuclear DVCS process, is the DVCS off a bound nucleon in a nucleus
as represented in Fig. \ref{fig:InCohDiag} for an helium-4 target. The remnants of 
the nucleus ($X$) contain only the missing three nucleons. 
The theory for incoherent DVCS on the nucleon is largely based on the free proton theory
already reviewed widely in the literature~\cite{Diehl:2003ny,Belitsky:2005qn,Guidal:2013rya}. 
Two important differences need to be accounted for however: the different initial state and the addition of 
FSIs. In the initial state, the intrinsic Fermi motion of the nucleons in the nucleus 
leads to an uncertainty on the exact kinematics of the reaction. Moreover, in general, the nucleon 
is in an off-shell state that is not exactly identical to its final state. In the final state, 
interactions between the outgoing nucleon from the DVCS reaction and the remnants of the nuclear 
target are possible. The latter leads to contamination from other channels; in particular, 
charge exchange processes can lead to a large contribution from such background reactions. 

Since DVCS is a 
process selected using tight exclusivity constraints, some of the initial and final-state effects are
automatically mitigated. Selection criterion on missing energy and momentum are performed,
constraining the range of initial Fermi motion and FSIs possible. However, no theoretical
calculation is available to correct for the reminder of these effects yet. Modern calculations exist 
for such effects in deep inelastic scattering \cite{Cosyn:2017ekf} and quasi-elastic 
scattering \cite{Ethier:2014bua}, and we can expect them to be extended to the DVCS 
process as more data become available. Another avenue of progress
on this topic will be the use of experimental techniques like tagging. This process
can help to control both initial and final state effects by detecting the nuclear remnant.
In the tagged process the target breaks in two, thus measuring the nuclear remnant provides 
information about the initial state of the struck nucleon, while a backward fragment also 
limits significantly the probability of FSIs.

\section{Past Nuclear DVCS Measurements}

\begin{figure}[tbp!]
\center
	\includegraphics[width=10.5cm]{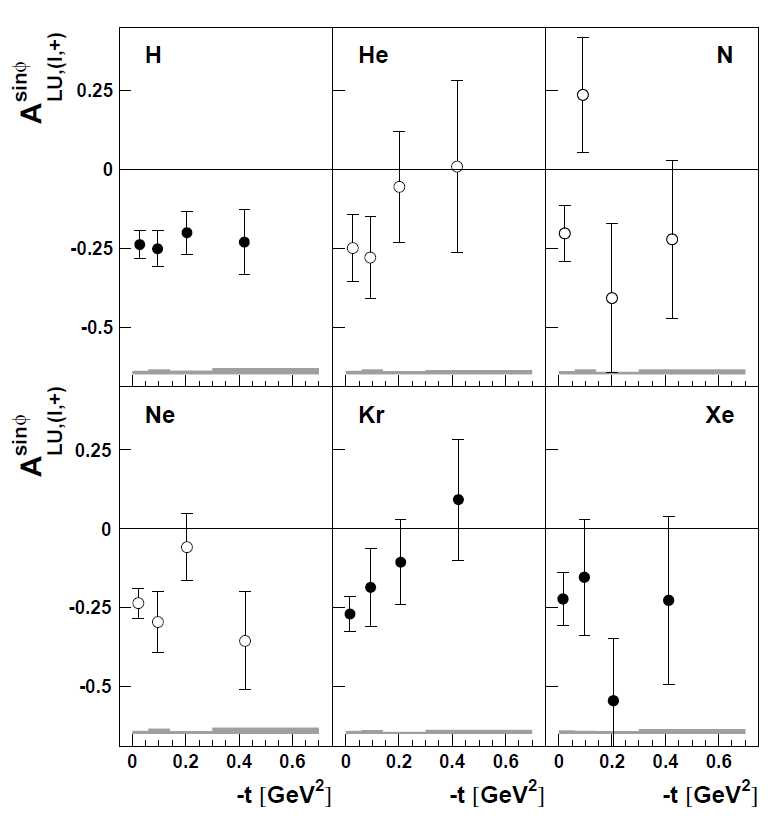}
	\caption{The $\sin(\phi)$ moment of the BSA as a function of $-t$ measured by HERMES
	for a series of nuclei \cite{Airapetian:2009cga}. The gray bands represent the systematic 
        uncertainties.}
\label{fig:HERMES1}
\end{figure}

\begin{figure}[tbp!]
\center
\includegraphics[width=10cm]{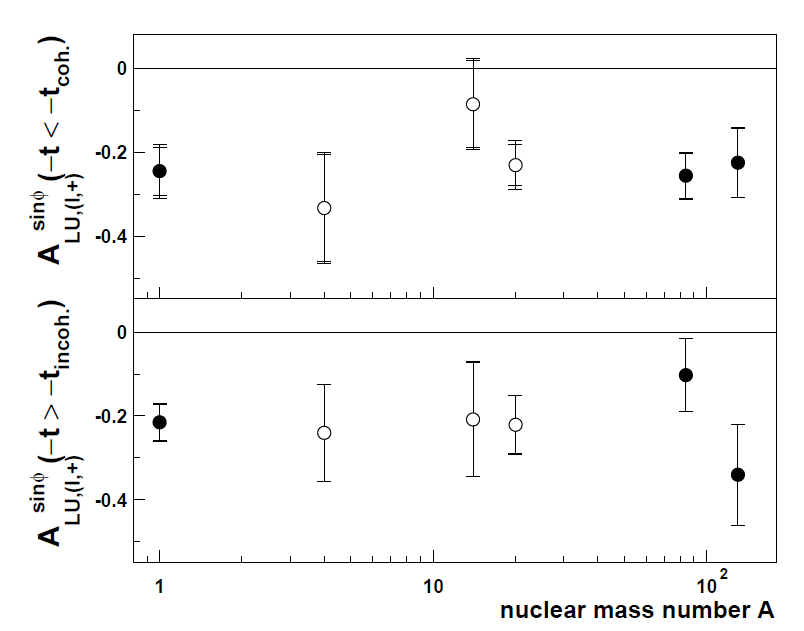}
\caption{The $\sin(\phi)$ moment of the BSA at low and high $-t$ as a function of $A$ measured by HERMES
	\cite{Airapetian:2009cga}. The inner error bars represent the statistical uncertainty, while 
the outer represent the quadratic sum of the statistical and systematic uncertainties.}
\label{fig:HERMES2}
\end{figure}

The first measurement of nuclear DVCS was performed by the HERMES 
Collaboration~\cite{Airapetian:2009cga}. This experiment covered an array of 
nuclear targets and looked at the $A$ dependence of the
BSA signal. Their main results, reproduced in Figs. \ref{fig:HERMES1} and \ref{fig:HERMES2}, suffer
from large uncertainties, which makes them consistent with the free proton data and prevents us to
reach strong conclusions about possible nuclear effects. Yet, in the coherent DVCS case a rather
strong effect was expected, leading to an apparent conflict between the HERMES results and theoretical
expectations. 

A characteristic of the HERMES measurement and how it was obtained from data can however
explain the discrepancy with theoretical expectations~\cite{Guzey:2003jh,Airapetian:2009cga}.
The main point being that the DVCS process is not fully detected and the scattered target
is instead reconstructed through a missing mass measurement of the other reaction products. The 
issue with this method is that the detector resolution is not good enough to separate the 
coherent and incoherent channels properly. 
Thus, the results are labeled "coherent enriched" and "incoherent enriched" at low and high 
$-t$, respectively. This label is based on the assumption that the very different behavior of the
cross sections of the two channels in $t$ will lead to a clear differentiation. However, the
results in Fig. \ref{fig:HERMES2} show similar behaviors in both sectors of $t$, which  
challenges this assumption and could explain the tension between theory and experiment. 

Altogether, large error bars and the impossibility to properly separate the coherent and 
incoherent channels have strongly impaired the interpretation of the measurement 
and the conclusions that can be
obtained from it. The CLAS experiment presented here has profited largely from this
result and was designed specifically to solve these two issues of low statistics and exclusivity.

\section{The CLAS Nuclear DVCS Experimental Setup}

The CLAS nuclear DVCS experiment had as its main objectives to explore coherent DVCS on helium-4, to assess if
the predicted BSA increase could be observed, and to extract the helium-4 GPD. In order to
perform this measurement however, several instrumentation challenges needed to be resolved. First, to
measure the scattered electron and the small angle photon from DVCS, we used CLAS in its 
DVCS setup, $i.e.$ with the addition of a forward angle calorimeter and a 5-T solenoid magnet. Second, a
radial time projection chamber (RTPC) was installed to measure the
helium recoils and thus ensure the exclusivity of the process in the coherent channel. In this section, we 
will review the important elements of this detection setup. 

\subsection{The CEBAF Large Acceptance Spectrometer (CLAS)}

\begin{figure}[tbp!]
\center
\includegraphics[width=8cm]{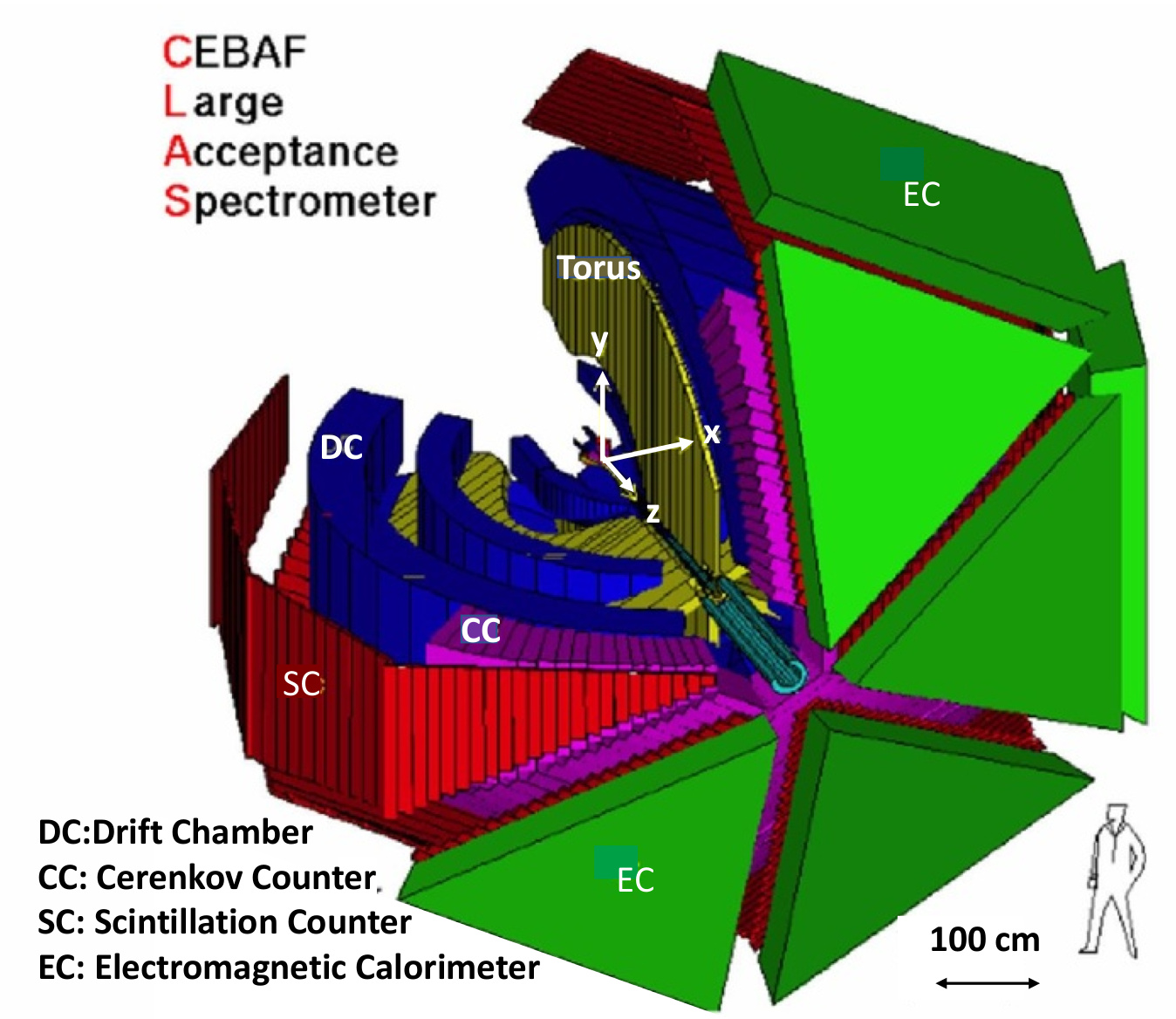}
	\caption{View of the CLAS detector setup.}
\label{fig:CLAS}
\end{figure}

The CLAS \cite{Mecking:2003zu} spectrometer was installed in Hall~B of Jefferson Lab (JLab) continuous electron beam 
accelerator facility (CEBAF). This detector was specifically designed to study the multi-particles 
final states that cannot be observed conveniently with multi-arm spectrometers. It was naturally 
well-suited for measuring DVCS, and several DVCS experiments were successfully conducted 
before this experiment using multiple different configurations. CLAS was composed of six identical 
sectors separated by the coils of a toroidal magnet, with each sector made of four detectors 
as shown in Fig. \ref{fig:CLAS}. Three regions of drift chambers \cite{Mestayer:2000we} 
were placed between the torus magnet to reconstruct the charged particles' tracks and 
calculate their momentum. An array of scintillators was placed behind the drift chambers to measure 
the precise time-of-flight for each track
\cite{Smith:1999ii}. These detectors covered the polar angle from 8 to 142 degrees. 
In the forward region, from 8 to 45 degrees, these detectors were complemented with 
Cerenkov counters \cite{Adams:2001kk} and electromagnetic calorimeters \cite{Amarian:2001zs},
important for electron identification and photon detection. 

Altogether, CLAS provided a large acceptance for momenta 
starting at 200~MeV. The nuclear DVCS experiment took place from October to December 2009
at an electron beam energy of 6.064 GeV, with the beam intensity varying
between 120 and 150 nA. This beam, on the helium-4 target pressurized between 5 and 6 atm,
corresponds to luminosities in the range of $1$ to $1.2 \times 10^{34}$ cm$^{-2}$s$^{-1}$.
During the experiment, the data acquisition operated at a rate of about 3 kHz with about 70\% live-time
using an inclusive electron trigger. 

\subsection{Adaptations for DVCS}

The CLAS Collaboration has established a specific setup to measure the typically 
small angle photons of the DVCS process. This setup is composed of an inner 
calorimeter and a solenoid and has been employed for numerous DVCS measurements on proton 
targets \cite{Seder:2014cdc,Jo:2015ema,HirlingerSaylor:2018bnu}.

\begin{figure}[tbp!]
\center
\includegraphics[width=8cm]{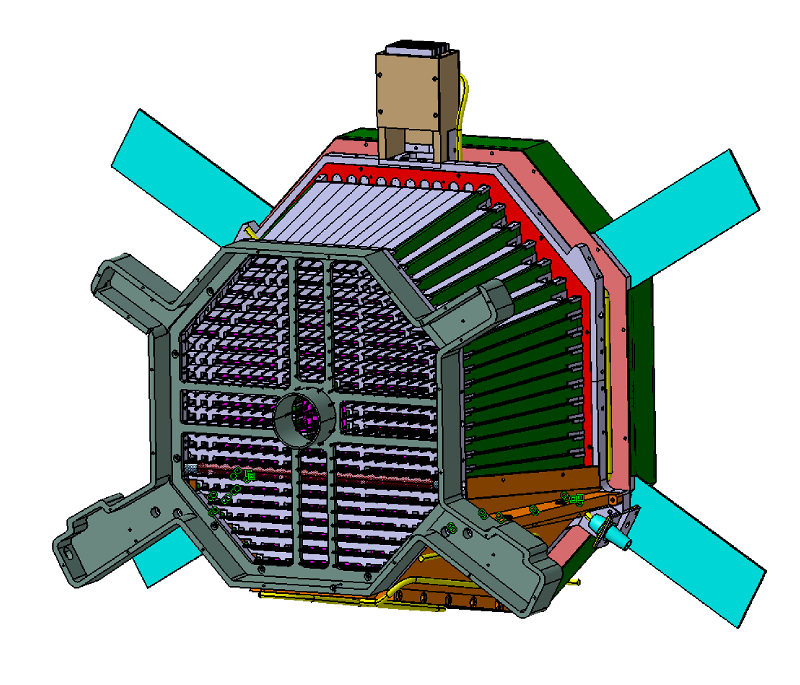}
	\caption{Representation of the inner calorimeter (IC) of CLAS. The crystals that compose the
	sensitive part of the detector are represented in purple.}
\label{fig:IC}
\end{figure}

\begin{figure}[tbp!]
\center
\includegraphics[width=14cm]{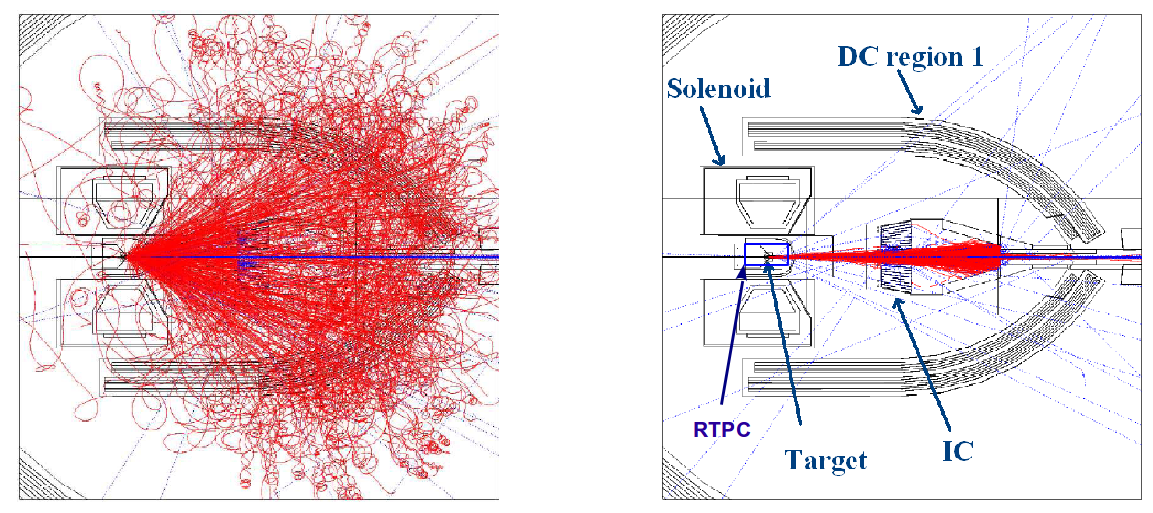}
\caption{Representation of the center of CLAS with the beam background in red with and without
	the solenoid field activated, right and left, respectively.}
\label{fig:Solenoid}
\end{figure}

The inner calorimeter, illustrated in Fig. \ref{fig:IC}, is a homogeneous 
calorimeter composed of 424 lead tungstate 
(PbWO) crystals read out by $5 \times 5$ mm$^2$ avalanche photo-diodes (APDs). 
It covers angles from 4 to 15 degrees. However, placing a detector at such small angles makes it 
particularly sensitive to the low energy Moller electrons scattered from the target.
To protect the calorimeter from this background, a 5~T solenoid was 
placed around the target to form
a magnetic shield. Thanks to this field, low energy charged particles (particularly 
electrons) curled around the beamline 
and never made it to the calorimeter or other CLAS detectors as illustrated 
by the simulation results presented 
in Fig. \ref{fig:Solenoid}. This allows to run much higher luminosity experiments,
a necessity for low rate processes like DVCS.

\subsection{The Radial Time Projection Chamber}

The recoil helium nuclei from coherent DVCS are mostly emitted between 150 
and 200~MeV at the beam energy of 6~GeV. Therefore, a specific detector was 
needed to detect them. To design the present setup, inspiration was drawn from the 
BONUS setup that also used a GEM-based RTPC~\cite{Fenker:2008zz} in CLAS to 
detect slow protons coming out of a deuterium target~\cite{Baillie:2011za}. In such an RTPC the 
ionization electrons drift toward large radii 
rather than toward the endcaps, as is more traditional in time 
projection chambers. This design allows to reduce significantly the drift time and reduce the 
amount of pile-up from accidental events. The RTPC design, its operation, calibration and 
the track reconstruction have been described in more 
details elsewhere~\cite{Dupre:2017upj}. Here a summary of key elements is provided.

\begin{figure}[tbp!]
\center
\includegraphics[width=10cm]{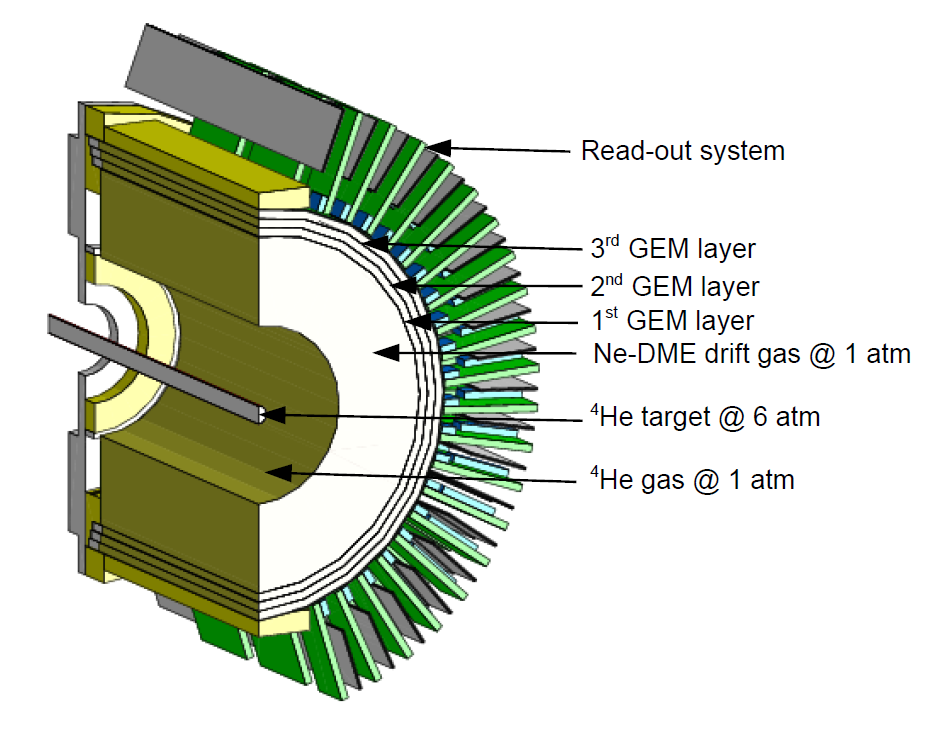}
\caption{Cut view of the RTPC.}
\label{fig:RTPCGlobal}
\end{figure}

In order to detect the recoil helium nuclei from a DVCS reaction, we first need 
to ensure that it will come out of the target. For this, we used a light straw 
target made of a thin kapton wall of 27 $\mu$m filled with helium at 6 atm 
pressure. The entrance and exit windows are thin aluminum foils and an helium 
bag was placed downstream of the target to avoid interaction with air in the gap 
between the target and the beamline vacuum. The cylindrical 
chamber surrounds the target as illustrated in Fig.~\ref{fig:RTPCGlobal}. Here we 
list the elements composing it based on their radii:
\begin{itemize}
	\item Up to a radius of 3 mm the pressurized helium target.
	\item From 3 to 20 mm a keep-out zone filled with 1 atm of helium to 
		minimize the production of secondaries. 
	\item At 20 mm a grounded foil made of 4 $\mu$m aluminized Mylar to 
		isolate the chamber from the beamline 
		region and collect charges. It also serves to separate the gas regions.
	\item From 20 to 30 mm a dead zone
		filled with the drift gas to separate the ground from the cathode.
	\item At 30 mm the cathode foil made of 4 $\mu$m aluminized Mylar.
	\item From 30 to 60 mm the drift region filled with the drift gas, a mix of 
                neon and dimethyl ether (DME) in an 80/20 proportion.
	\item From 60 to 69 mm the amplification regions, filled with drift gas, 
		with GEM foils placed at 60, 63 and 66 mm.
	\item At 69 mm the collection pads connected to the preamplifers placed 
		directly outside the chamber.
\end{itemize}

The time-to-position calibration of the detector has been performed with a 
dependence on $z$, the position along the beamline axis, due to 
variations in the magnetic fields. To perform this calibration we took  
dedicated data at 1.2 GeV beam energy. In 
this data set, we were able to select elastic events, for which the kinematics
of the helium recoil can be calculated from the electron kinematics and 
directly compared to the measurement in the RTPC. This comparison helped to 
map the correspondence between time and position in the chamber and determine 
the drift path of electrons. A more detailed description of the calibration process 
is available in Ref.~\cite{Dupre:2017upj}.

\section{DVCS Event Selection}

\subsection{Particle Identification}

The scattered electrons were detected with the baseline CLAS detectors. The drift chamber measured the kinematics
of the electron and the signal measured in both the Cerenkov counter and electromagnetic calorimeter
provided the identification. A signal of good quality was also required in the 
time-of-flight system, which served as a time reference for all detectors. 
Protons were detected with the baseline CLAS detectors as well, the drift chamber measured the kinematics
of the proton and the time-of-flight system ensured its identification. Several 
fiducial cuts are applied to ensure that particles did not go through part of the inner calorimeter 
or the solenoid, as well as to reject the edges of the detectors, where their efficiency is rapidly 
decreasing. Kinematic corrections are also applied to the electrons and protons to correct for energy loss
and biases in calibration, which are at the subpercent level except for protons below 500 MeV for which they
go up to 10\% at the detection limit of 200 MeV.

The photons from DVCS are mainly detected with the inner calorimeter. No specific 
identification cuts were used in this detector as large energy deposit was dominantly 
from electrons and photons, which could not be separated reliably. However, the detection 
of an electron at large angle in CLAS highly suppressed the number of electrons 
in the calorimeter; moreover, the exclusivity cuts used later in the analysis further 
this suppression. Left-over accidentals were accounted for in the background subtraction 
described below. The inner calorimeter was calibrated through a series of steps, 
involving the reconstruction of $\pi^0$ from their decay into two photons. Calibration was obtained 
with an iterative process to adjust each crystal gain to obtain the most accurate $\pi^0$ mass. A
global calibration of the calorimeter was also performed to account for incident angle, energy and 
time dependent effects.

The helium-4 nuclei were detected with the RTPC using a series of constraints on the quality
of the track reconstruction. As the chamber was operated at low gain and had very low efficiency 
for protons, we did not apply further identification cuts for the helium-4 nuclei 
detection~\cite{Dupre:2017upj}.

Finally, we selected events that contain a single electron, a high energy photon ($E>2$ GeV) and
either a helium or a proton. We applied a selection cut on the two charged particles 
to ensure they originated from the same vertex inside the target, thus rejecting accidentals and
events from the target windows.
Moreover, since we are aiming to study deep processes occurring at the partonic level, we selected
$Q^2>1$~GeV$^2$. Also, the transferred momentum squared to the recoil $^{4}He$ was bound by
   a minimum value based on basic energy-momentum conservation:
\begin{equation}
   t_{min} = - Q^{2} \frac{2(1- x_{A})(1 - \sqrt{1 + \epsilon ^{2}}) + \epsilon 
   ^{2}}{4 x_{A}(1-x_{A})+ \epsilon ^{2}},\\
\end{equation}
where $\epsilon ^{2} = \frac{4M^{2}_{^4\!He}x^{2}_{A}}{Q^{2}}$. For incoherent DVCS, 
we used a similar cut where $x_A$ is replaced by $x$ and $M_{^4\!He}$ by $M_{p}$. 

\subsection{Exclusive Photo-Production Selection}

In principle, 
a selection based only on the missing energy of the system would be
enough to guarantee the 
exclusivity of the process. However, in our experiment, where particles 
were detected at very different energies and with very different detector 
resolutions, this method was not sufficient. For instance the momentum 
of the helium nuclei is negligible in the missing energy observable, thus
this valuable information has no impact on a selection using this observable only.
To address this issue, we constrained the selection of our exclusive events by using seven 
variables selected to optimize the use of all the detector information available. 
The seven variables are defined as follows for coherent 
DVCS case (replace helium by proton for the incoherent case):
\begin{itemize}
	\item Co-planarity ($\Delta \phi$) of the virtual photon, the real photon and
		the recoil helium;
	\item Missing energy of the complete final state;
	\item Missing mass of the complete final state;
	\item Missing transverse momentum of the complete final state;
	\item Missing mass of the electron-helium system;
	\item Missing mass of the electron-photon system;
	\item Co-linearity ($\theta$) of the measured photon with the missing momentum of the 
		electron-helium system.
\end{itemize}

\begin{figure}[tbp!]
\center
\includegraphics[trim=10 10 10 5,clip,width=16.5cm]{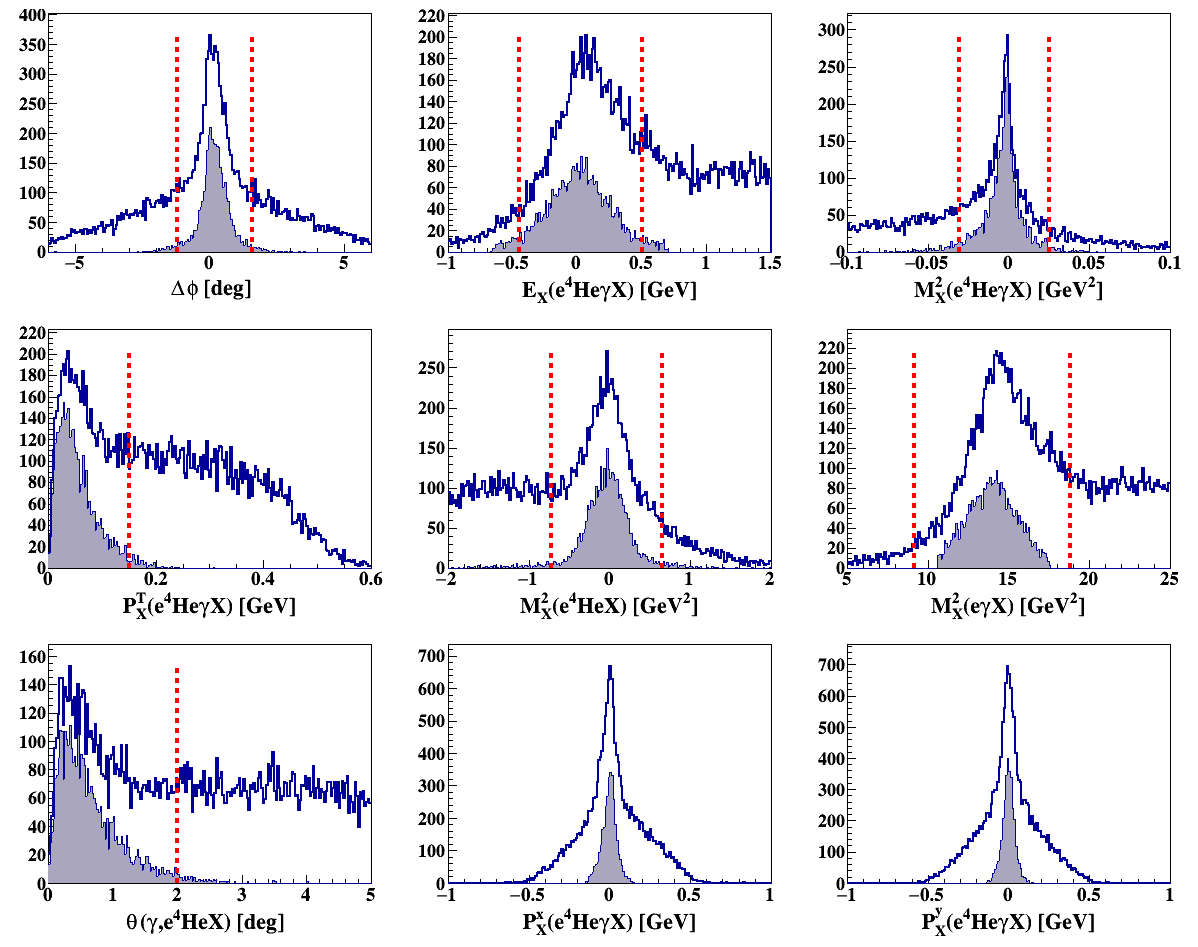}
        \caption{Distributions of the coherent photon production events before 
        (blue) and after (black line filled
	in gray) the exclusivity cuts used to select coherent DVCS represented by the 
	red dashed lines. The histograms are shown as a function of the 
	seven variables used for the exclusivity selection described in the text, 
	plus the missing $P_x$ and $P_y$ components, in order left to right and top to bottom. 
	}
\label{fig:CohExcCuts}
\end{figure}

\begin{figure}[tbp!]
\center
\includegraphics[trim=10 10 10 
   5,clip,width=16.5cm]{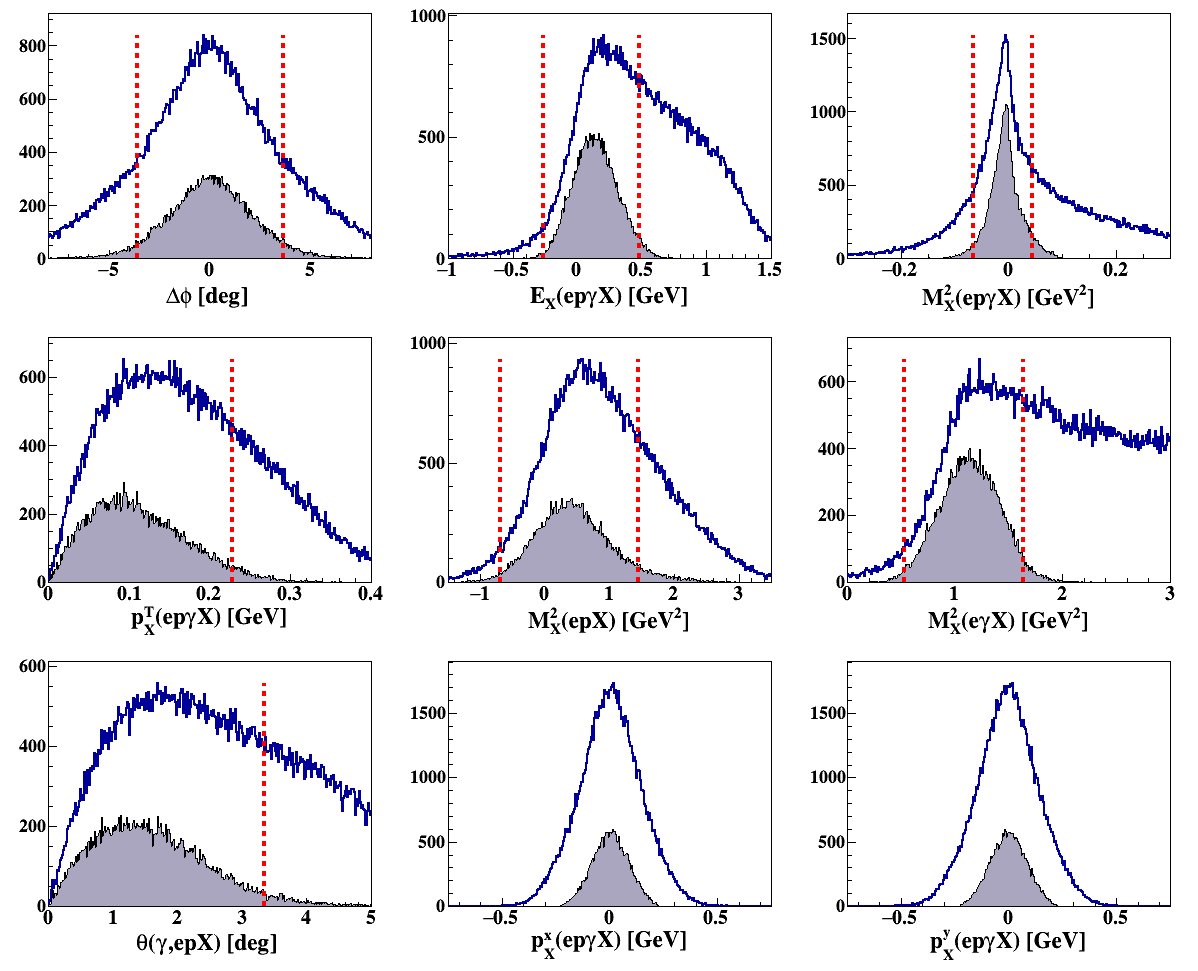}
\caption{Distributions of the incoherent photon production events before (blue) 
   and after (black line filled
	in gray) the exclusivity cuts used to select coherent DVCS represented by the 
	red dashed lines. The histograms are shown as a function of the 
	seven variables used for the exclusivity selection described in the text, 
	plus the missing $P_x$ and $P_y$ components, in order left to right and top to bottom.}
\label{fig:IncExcCuts}
\end{figure}

In the analysis, we applied selection cuts based on a fit of the exclusive peak at 3$\sigma$ around 
the mean value for each variable. This systematic method helps to avoid any bias 
in the selection of the events. The selection of coherent DVCS with these variables is illustrated in 
Fig. \ref{fig:CohExcCuts}. We note on these distributions only a few minor anomalies, where the 
distributions have some asymmetries. These are linked with the detector resolution, which impact
some of the kinematic variables non-linearly. The selection of incoherent DVCS is presented in 
Fig. \ref{fig:IncExcCuts}, with two main differences: wider distributions and larger
offset from the nominal expectations. The wider distributions are mainly attributed to the effect of 
Fermi motion, but simulations have shown that this effect is not strong enough to fully 
reproduce the distribution widths and FSIs must play a role as well. 
The offsets of some distributions are caused by slight detector 
misalignment between CLAS sectors and are within the levels obtained with free proton 
targets~\cite{HirlingerSaylor:2018bnu} to which they can be directly compared. 

\subsection{Background Subtraction}

\begin{figure}[p]
\center
\includegraphics[trim=70 15 70 70,clip,width=6.3cm]{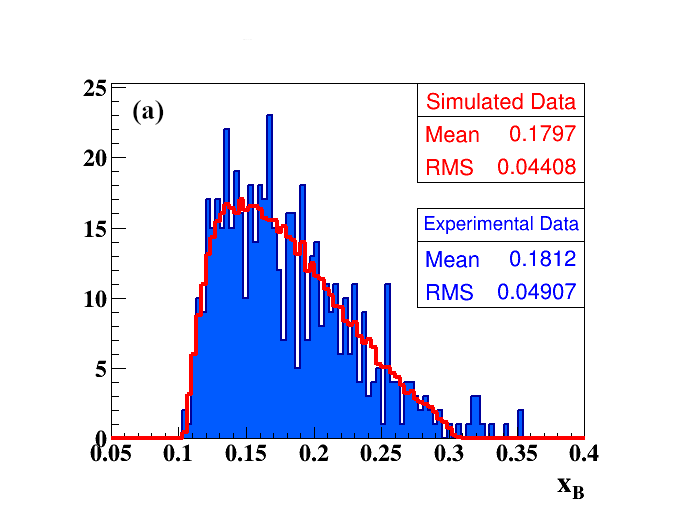}
\includegraphics[trim=70 15 70 70,clip,width=6.3cm]{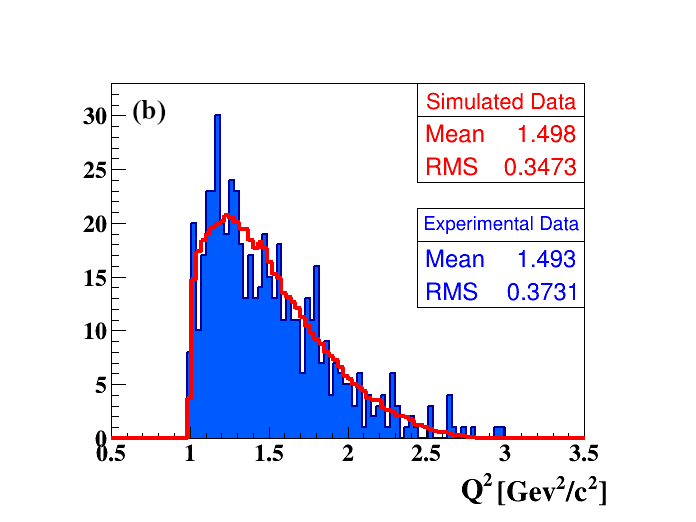}
\includegraphics[trim=70 15 70 70,clip,width=6.3cm]{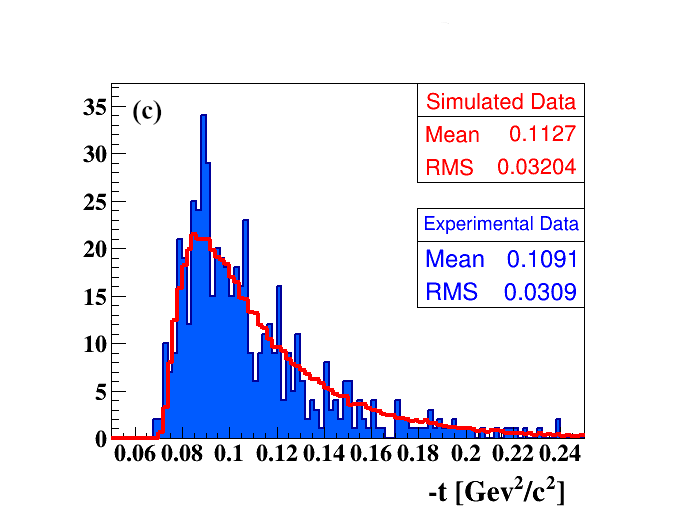}
\includegraphics[trim=70 15 70 70,clip,width=6.3cm]{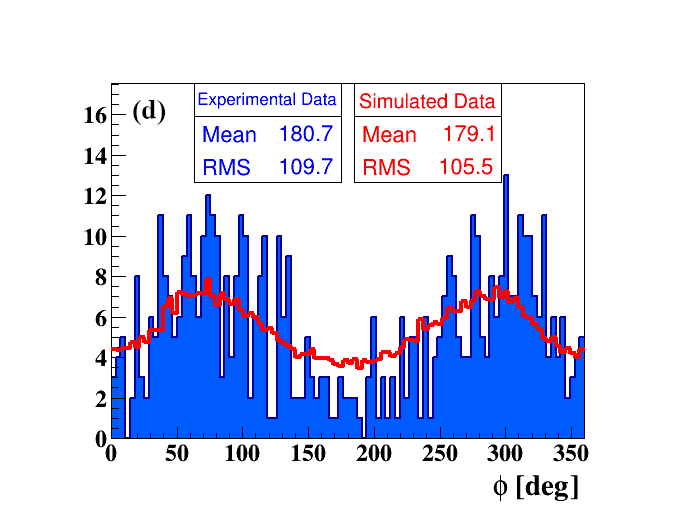}
	\caption{The measured (filled blue) and simulated (red) distributions of
	coherent exclusive $\pi^0$ production as a function of $x$ (panel a), 
        $Q^2$ (panel b), $-t$ (panel c) and $\phi$ (panel d).}
\label{fig:CohPi0Simul}
\end{figure}

\begin{figure}[p]
\center
\includegraphics[trim=70 15 70 70,clip,width=6.3cm]{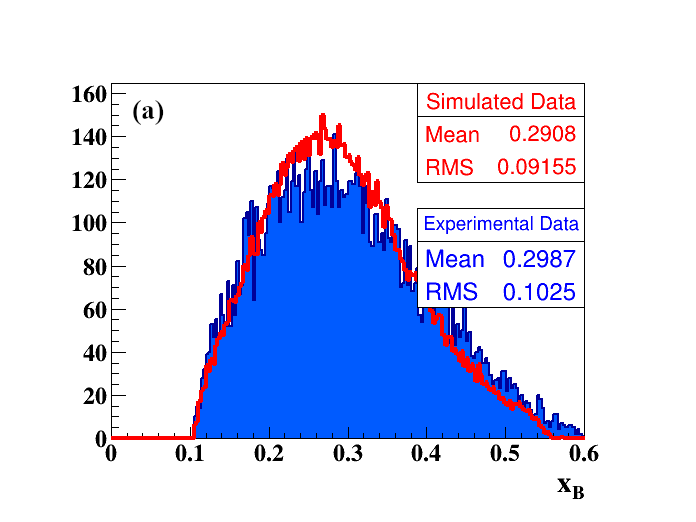}
\includegraphics[trim=70 15 70 70,clip,width=6.3cm]{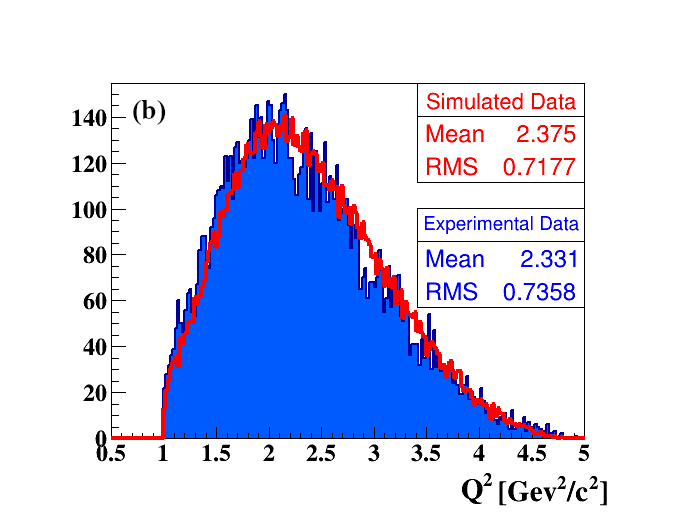}
\includegraphics[trim=70 15 70 70,clip,width=6.3cm]{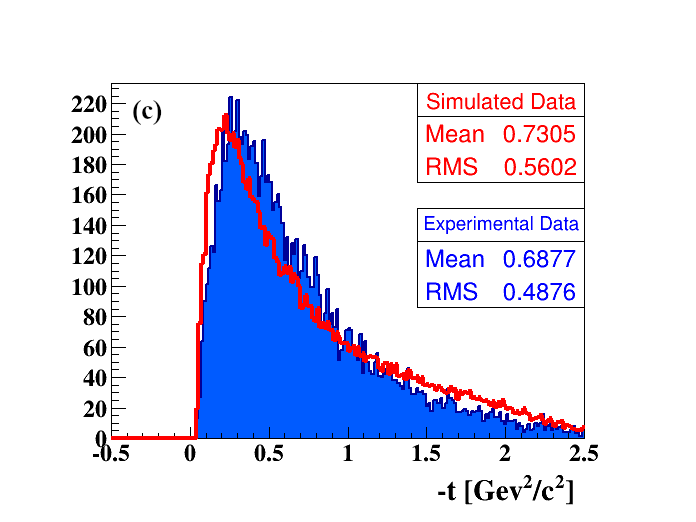}
\includegraphics[trim=70 15 70 70,clip,width=6.3cm]{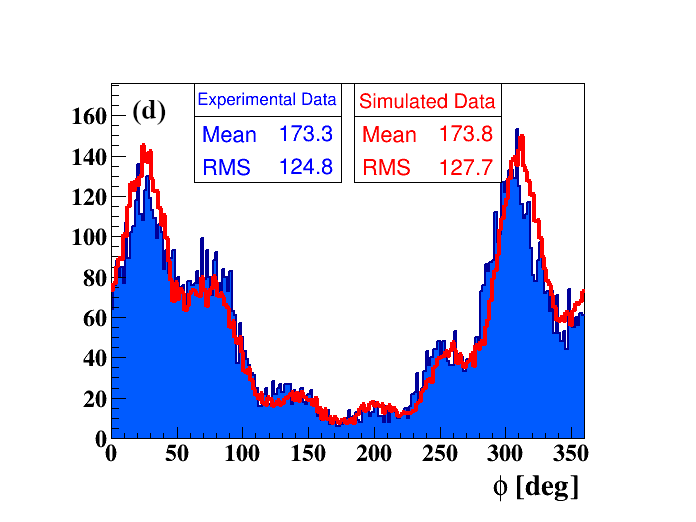}
	\caption{The measured (filled blue) and simulated (red) distributions of
	incoherent exclusive $\pi^0$ production as a function of $x$ (panel a), 
        $Q^2$ (panel b), $-t$ (panel c) and $\phi$ (panel d).}
\label{fig:InCohPi0Simul}
\end{figure}

The main signal contamination comes 
from the exclusive production of a $\pi^0$, the final state of which is very similar to DVCS with 
only an extra photon. In such an event, if one of the photons is produced at low energy, it is 
easy to confuse this process with single photon production. In order to estimate the contribution from
this channel in the data, we measured the exclusive $\pi^0$ production in the same way as DVCS, with 
a series of exclusivity cuts, completed by a selection cut on the invariant mass of the two photons
to match the $\pi^0$ mass. The events obtained for the coherent and incoherent
channels are shown in Figs. \ref{fig:CohPi0Simul} and \ref{fig:InCohPi0Simul}, respectively. Using 
this sample, we developed an event generator and adjusted it to the data. The result of which 
is shown with the red histograms of Figs. \ref{fig:CohPi0Simul} and 
\ref{fig:InCohPi0Simul}. To correct the experimental data, we then estimated the number of single photon 
events coming from the exclusive $\pi^0$ production as:
\begin{equation}
	N_{1\gamma,\pi^0}^{Exp} = \frac{N_{1\gamma,\pi^0}^{Sim}}{N_{2\gamma,\pi^0}^{Sim}} \times N_{2\gamma,\pi^0}^{Exp},\\
\end{equation}
where $N_{1\gamma,\pi^0}^{Sim}$ is the number of simulated exclusive $\pi^0$ mistaken for DVCS events,
$N_{2\gamma,\pi^0}^{Sim}$ the number of simulated exclusive $\pi^0$ fully reconstructed and $N_{2\gamma,\pi^0}^{Exp}$
the number of experimentally measured exclusive $\pi^0$. This number was then subtracted from
the experimentally measured number of DVCS events ($N_{DVCS}^{Exp}$) to get the corrected result: 
\begin{equation}
	N_{DVCS}^{Corr} = N_{DVCS}^{Exp} - N_{1\gamma,\pi^0}^{Exp}.\\
\end{equation}

\begin{figure}[tbp]
\includegraphics[width=8cm]{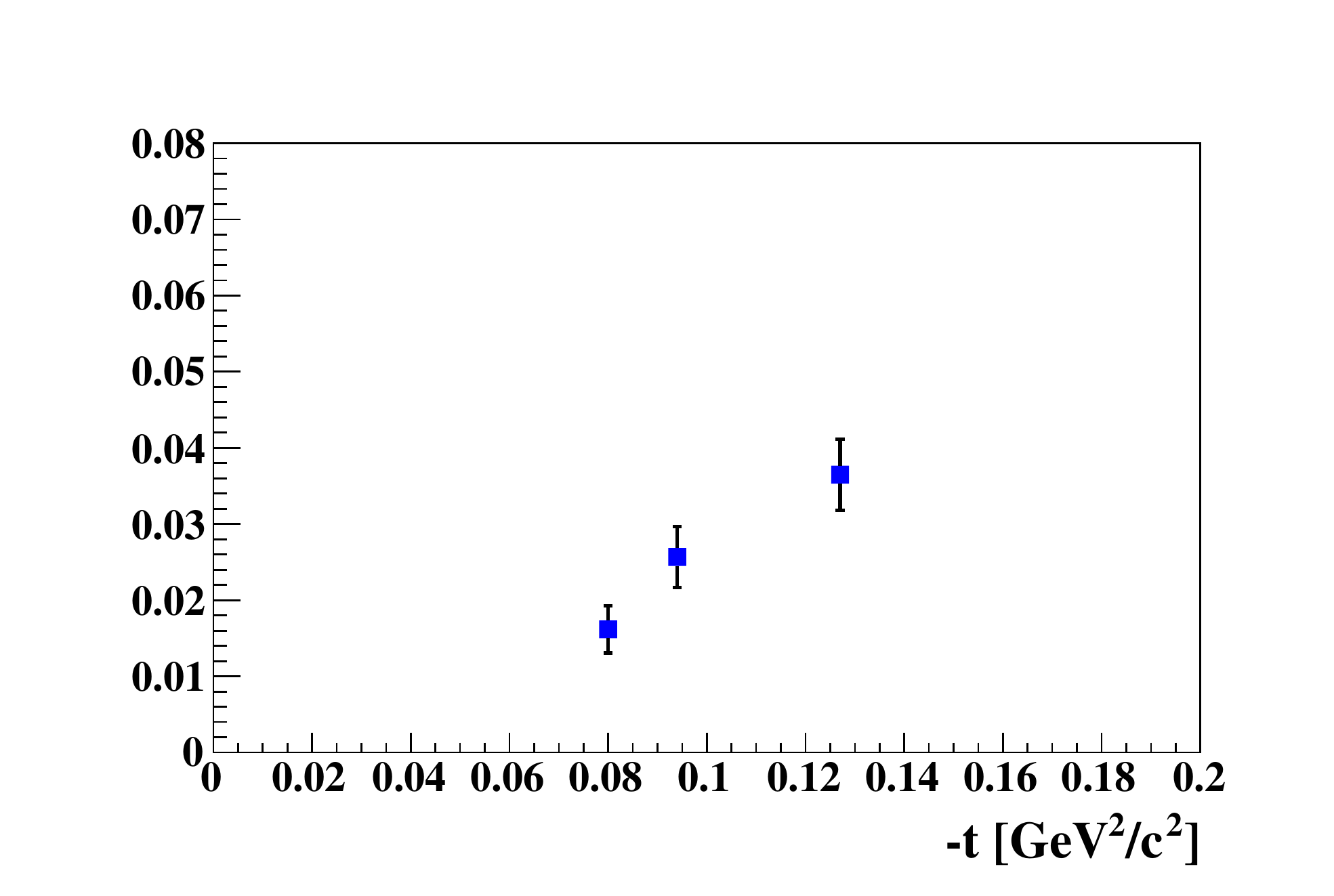}
\includegraphics[width=8cm]{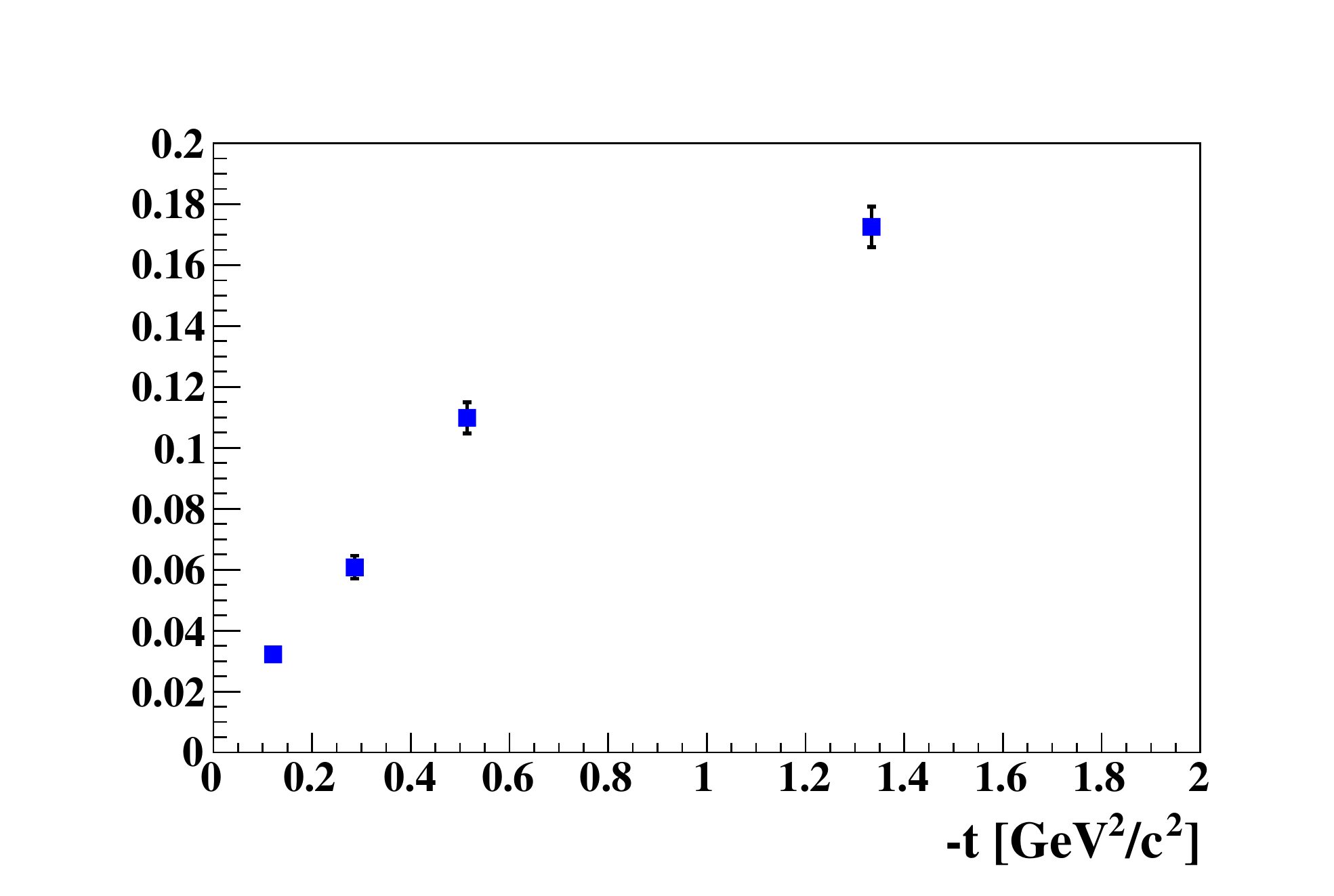}
\caption{The estimated coherent (left) and incoherent (right) $\pi^{0}$ 
contamination fraction in the DVCS events as a function of the 
transferred momentum squared -$t$ and integrated over the kinematic 
variables $Q^2$, $x_B$, and $\phi$.}
\label{fig:cont_yield}
\end{figure}

We show in Fig.~\ref{fig:cont_yield} the $\pi^0$ contamination for the $-t$ 
bins, where it varies the most from one bin to another. The study shows 2 to 
4\% contamination in the coherent channel and 3 to 17\% in the incoherent 
channel. After subtracting this contamination from the denominator of the 
asymmetry, we make no further correction to 
the DVCS BSA, \textit{i.e}., we assume the exclusive $\pi^0$ production has no such 
asymmetry in either the coherent or incoherent channels. Our own exclusive $\pi^0$ 
data rules out any BSA above approximately 10\%, a level which would have an 
insignificant effect on our results given the small amount of contamination.

The second important source of background comes from accidentals. Despite the many exclusivity cuts, it is 
possible to have particles from different events being combined and pass all the cuts to get 
into the data sample. To evaluate the number of such events, we inverted the vertex selection 
of the two charged 
particles of the process, electron and helium (or proton in the incoherent case), and requested that they
are separate. We found that 4.1\% of the coherent and 6.5\% of the incoherent samples were accidentals,
they are also subtracted from the denominator of the asymmetry. 

\subsection{Systematic Uncertainties}

To further evaluate the systematic uncertainty of the measurements, we performed several specialized studies. We 
evaluated the impact of changing the exclusivity selection cuts by varying them from 1 to 5 $\sigma$. 
We also evaluated the impact of changing the binning in $\phi$ on the extraction of the BSA
at 90 degrees. The beam polarization was measured using M\o{}ller scattering runs, the uncertainty was 
estimated based on the known precision of the dedicated apparatus and the spread 
of the measurements during the complete run period. We studied how different methods of
simulating the exclusive $\pi^0$ production affected the single-photon background and
further estimated how much bias could arise from an undetected BSA in the process. As 
radiative corrections are expected to be small
for this process, we did not apply them, but associated an uncertainty equal to their expected value.
These uncertainties are summarized in Tab. \ref{Table:systematic_uncertainties}, with their respective 
evaluated values. They are added quadratically to obtain the total systematic uncertainty presented in the results.

\begin{table}[tbp]
\begin{center}
	\begin{tabular}{|m{4cm}|m{2cm}<{\centering}|m{2.3cm}<{\centering}|m{3.7cm}<{\centering}|}
\hline
\bf Systematic source & \bf  Coherent channel  & \bf Incoherent channel & \bf Type of systematic 
error\\
\hline
Beam polarization &  3.5$\%$ &  3.5$\%$& Normalization\\
\hline
\hline
DVCS cuts & 8 $\%$ &  6 $\%$ & Bin to bin\\
\hline
Data binning & 5.1$\%$ & 7.1$\%$ & Bin to bin\\
\hline
$\pi^0$ subtraction &  0.6$\%$ &  2.0$\%$ & Bin to bin\\
\hline
Radiative corrections &  0.1$\%$ & 0.1$\%$ & Bin to bin\\
\hline
\hline
\textbf{Total bin to bin} &  \textbf{10.1}$\%$ &   \textbf{10.1}$\%$ & Bin to 
bin\\
\hline
\end{tabular}
\caption{The systematic uncertainties on the measured coherent and incoherent 
BSAs at $\phi = 90$~deg.}
\label{Table:systematic_uncertainties}
\end{center}
\end{table}

\begin{figure}[tbp!]
\center
\includegraphics[width=15cm]{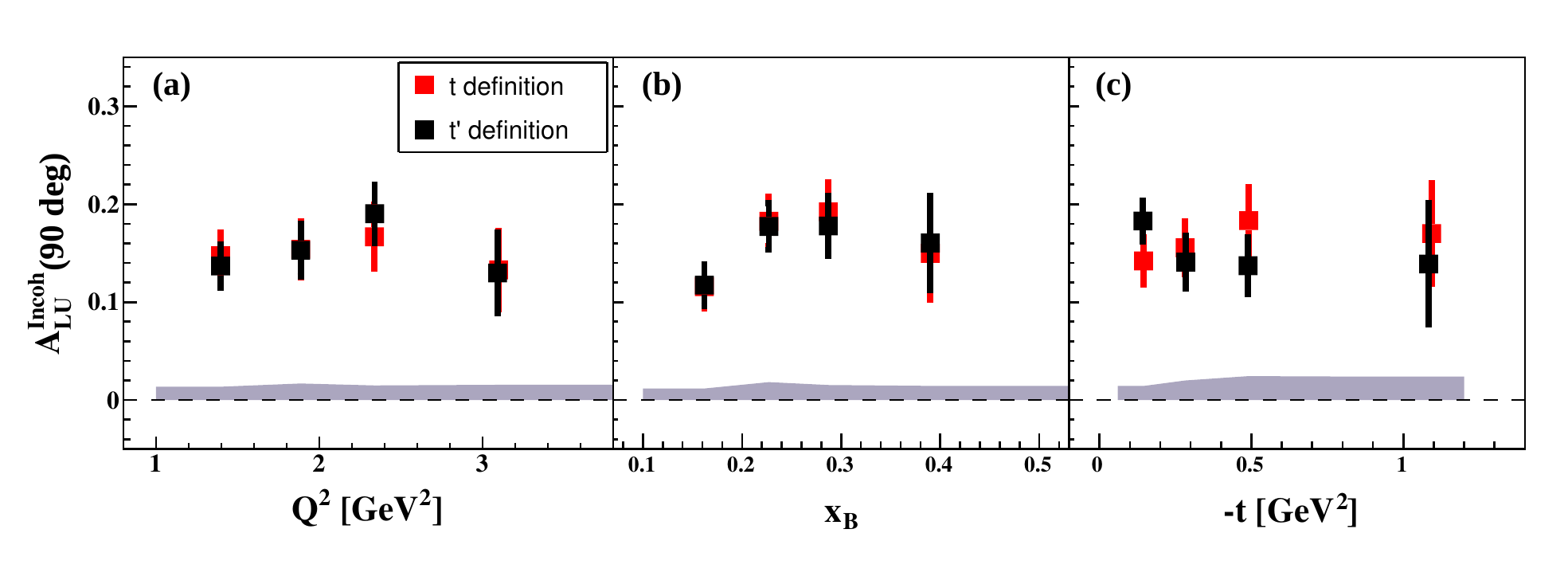}
	\caption{The BSA at 90 degrees ($A_{LU}^{Incoh} (90~\mathrm{deg})$) as a
	function of $Q^2$ (panel a), $x_{B}$ (panel b) and $-t$ (panel c), using the photon based $t$ definition (red)
	and the proton based $t^\prime$ definition (black).}
\label{fig:ttpComp}
\end{figure}

An extra problem that was studied is the best way to define $t$ in the incoherent channel, 
which is not completely straightforward. As can be seen in Fig. \ref{fig:InCohDiag}, 
we can either use $t$ or $t^\prime$ ($= (p - p^\prime)^2$). In principle, the two are 
identical, but experimentally we face some issues. The measurement of $t$ is less precise than
$t^\prime$ because it involves the photon rather than charged particles. However, the 
exact measurement of $t^\prime$ is impossible and one needs to assume a proton at rest
in the initial state to calculate $t^\prime$. As it is not obvious which solution is best,
we studied the difference between the two results by analyzing the data independently using the two 
definitions. We found no significant difference between them, as is 
illustrated in Fig. \ref{fig:ttpComp}. We use in the final results $t$ as it is based on the 
rigorous definition. Since the effect of resolution appears small and is partly accounted for
in the systematic uncertainty associated to the DVCS cuts, we decided not to associate an extra
systematic uncertainty based on this study.

\section{Results}

\subsection{Coherent DVCS}

\begin{figure}[bp!]
\center
\includegraphics[width=12cm]{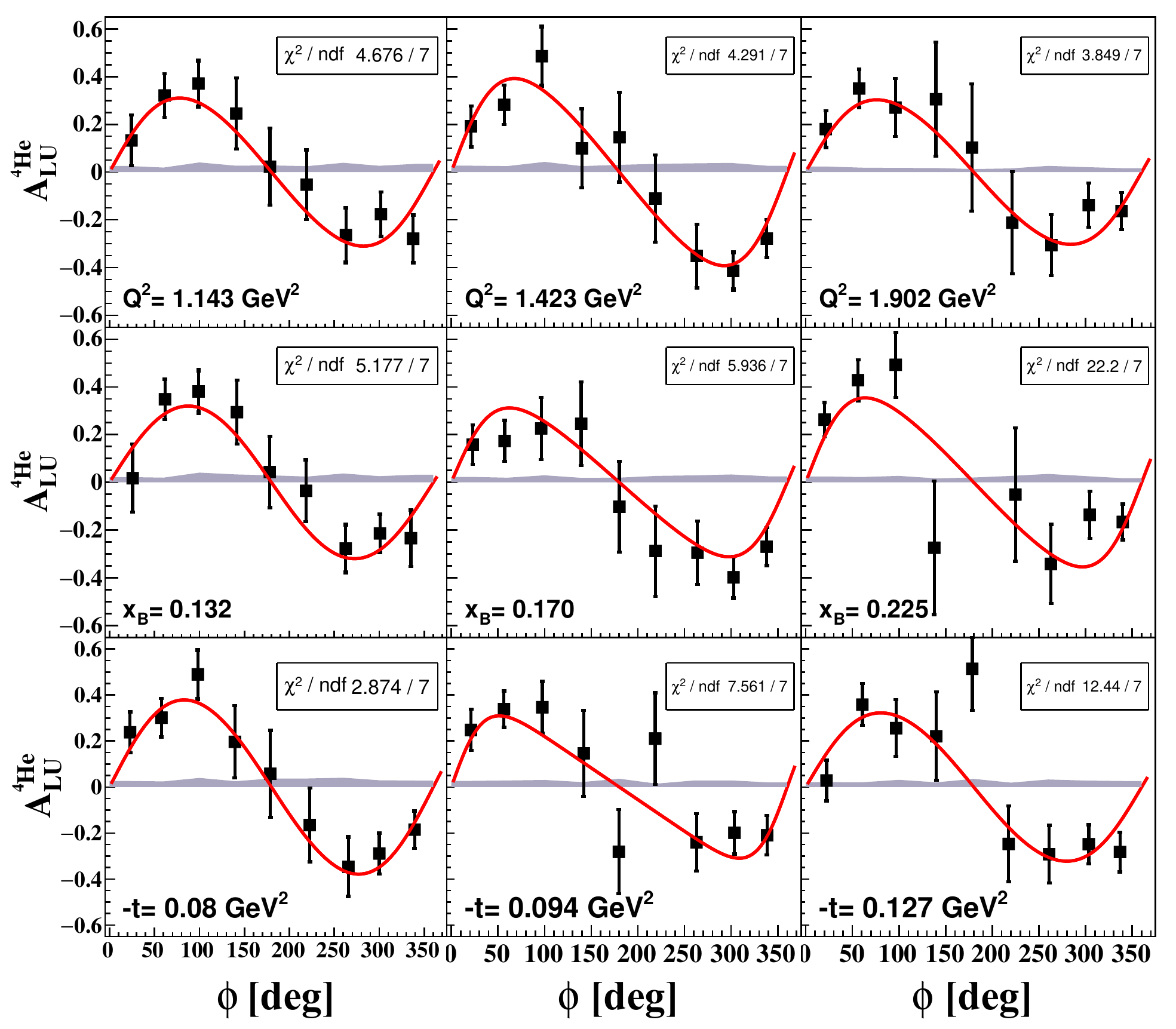}
	\caption{The BSA in the coherent exclusive photo-production off helium-4 as a 
	function of $\phi$ and $Q^2$ 
	(top panels), $x$ (middle panels) and $-t$ (lower panels). The error bars are  
	statistical and the gray bands represent the systematic uncertainties. The full red lines show
	the fit of the data with the form of Eq.~\ref{eq:A_LU-coh}.}
\label{fig:CohALUphi}
\end{figure}

In Fig. \ref{fig:CohALUphi}, we present the results for the BSA in the coherent DVCS channel. We 
observe the dominant sinusoidal component typical of the DVCS BSA, with an amplitude almost 
double that measured for the free proton \cite{Jo:2015ema}. This 
predicted feature of nuclear DVCS \cite{Guzey:2003jh} is observed here for the first time, due
to the fact that this measurement cleanly isolates the coherent DVCS process. The absence of 
this feature in the previous measurement by HERMES \cite{Airapetian:2009cga} and its clear 
observation here indicates that the recoil detection is necessary to isolate the effects of 
the coherent DVCS process from the incoherent background. 

\begin{figure}[tbp]
\center
\includegraphics[width=16cm]{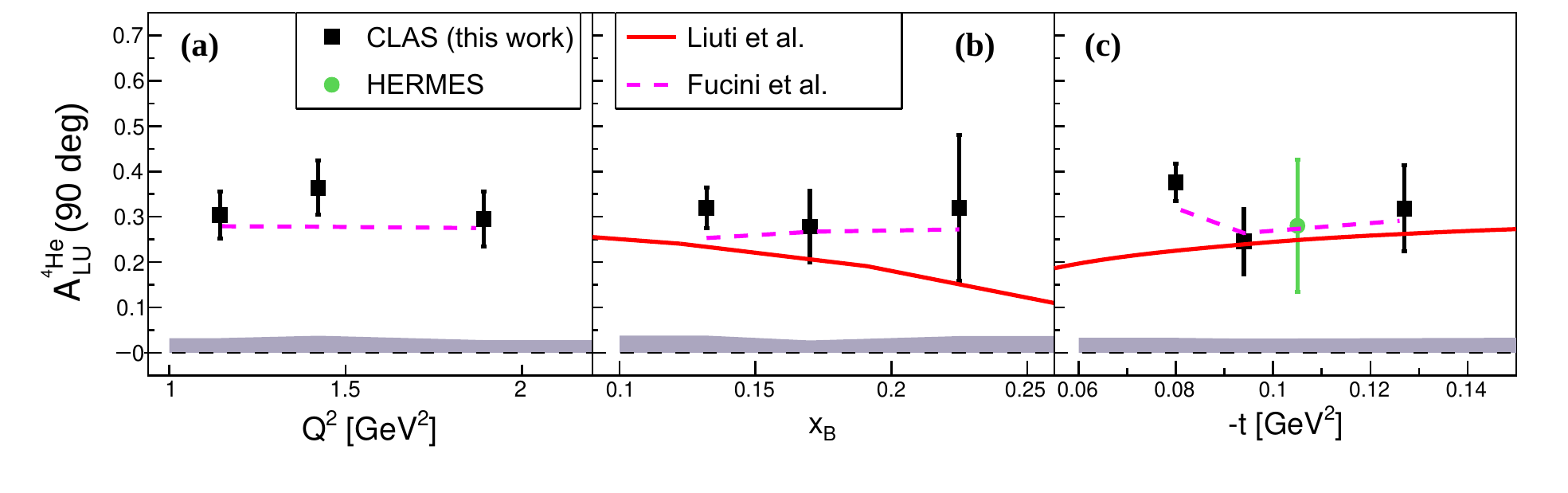}
\caption{The BSA at 90 degrees as a function of $Q^2$ (panel a), $x$ (panel b) and $-t$ (panel c).
	Our results are shown with black squares, HERMES results 
	with green circles~\cite{Airapetian:2009cga}. The theoretical prediction 
	by Liuti \textit{et al.}~\cite{Liuti:2005gi,GonzalezHernandez:2012jv} is shown 
	by the full blue line, while the calculation by Fucini 
	\textit{et al.}~\cite{Fucini:2018gso} is shown with the magenta dashed line.}
\label{fig:CohALU90}
\end{figure}

We show the extraction of the BSA at 90 degrees in Fig. \ref{fig:CohALU90} together with the 
past HERMES Collaboration results~\cite{Airapetian:2009cga}. Two models are compared to the 
data, they are both based on the hypothesis that the main nuclear effects are 
included by accounting for the nucleon off-shellness and the kinematics of nucleons in nuclei. The 
one by Liuti \textit{et al.}~\cite{Liuti:2005gi,GonzalezHernandez:2012jv} 
appears to undershoot the results systematically. However, the more recent and independent 
calculation by Fucini \textit{et al.}~\cite{Fucini:2018gso}, using similar principles but with a non-diagonal nuclear spectral 
function \cite{Viviani:2001wu} based on the AV18 nucleon-nucleon potential \cite{Wiringa:1994wb} 
and the UIV three-body forces \cite{Pudliner:1995wk}, has been able to reproduce the data very 
well. A factor in the difference is that the recent calculation by Fucini {\it et 
al.}~\cite{Fucini:2018gso} benefited from using the precise kinematics of each of the points presented 
in Appendix \ref{sec:fullresults}. Including this information appears to have a significant 
impact on some points, for instance the $-t$ distribution appears to have a peculiar structure 
that is well reproduced when using this information. 

\begin{figure}[tbp!]
\center
\includegraphics[width=14cm]{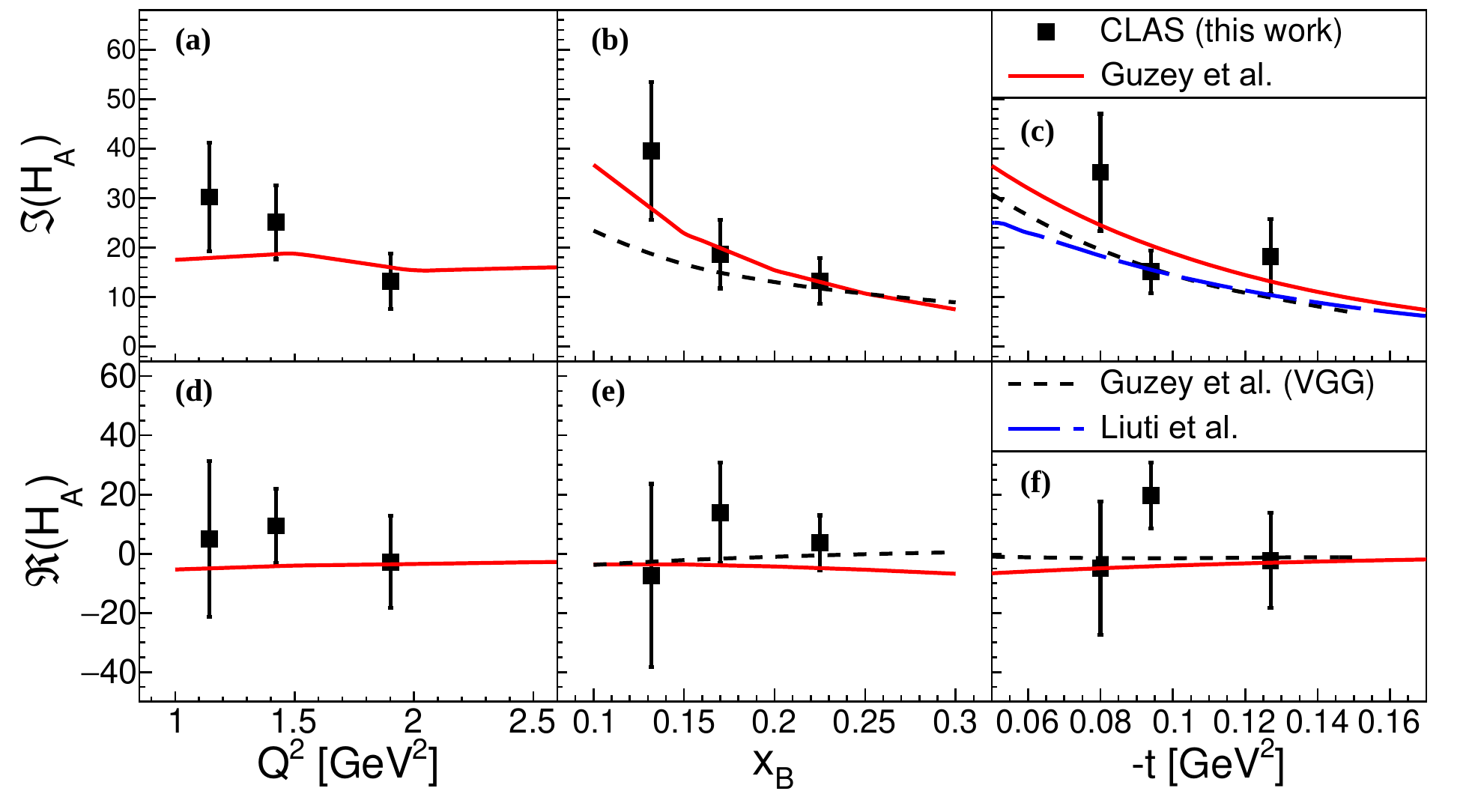}
	\caption{The imaginary part of the helium-4 CFF $\mathcal{H}_A$ is shown as 
        a function of $Q^2$ (panel a), $x$ (panel b) and 
	$-t$ (panel c). The real part of the helium-4 CFF $\mathcal{H}_A$ is shown as 
        a function of $Q^2$ (panel d), $x$ (panel e) and 
	$-t$ (panel f). The red full line is the theoretical calculation by 
	Guzey {\it et al.}~\cite{Guzey:2003jh,Guzey:2008th}, the black dashed line is the same calculation 
	using the VGG model as input \cite{Vanderhaeghen:1999xj,Guidal:2004nd}, and
	the blue long dashed line shows the predictions by Liuti {\it et 
	al.}~\cite{Liuti:2005gi,GonzalezHernandez:2012jv}.} 
\label{fig:CohCFF}
\end{figure}

One of the motivations for the choice of helium-4 for the coherent DVCS measurement was
a simplified extraction of the CFF $\mathcal{H}_A$ from the
data. To perform this step, we used the form from Eq. \ref{eq:A_LU-coh} to fit the data in Fig. 
\ref{fig:CohALUphi}. We present in Fig. \ref{fig:CohCFF} the extracted real and imaginary parts 
of the single CFF of the helium-4 nucleus. The results are rather encouraging. The two parts
of the CFF are constrained by data without the need for any model assumption. This capacity to obtain a
model independent result with such a limited data set offers a striking contrast with the
situation of the free proton fits \cite{Dupre:2016mai,Dupre:2017hfs}.

The CFF extraction allows us to compare the results to other theoretical calculations.
These are performed within the impulse approximation \cite{Guzey:2003jh,Guzey:2008th} 
and give the nuclear GPD directly from the proton and neutron GPDs. In 
Fig. \ref{fig:CohCFF}, we show two versions of this calculation, where two different nucleon GPD
models are used as input, compared with the calculation previously shown by Liuti 
{\it et al.}~\cite{Liuti:2005gi} with an updated nucleon model \cite{GonzalezHernandez:2012jv}.
We can see that the effect of changing the input nucleon GPD model
is of similar size or larger than the difference between the nuclear 
models. However, at the level of precision of the present data,
it is not possible to resolve which variant is best. This feature highlights the 
importance of the choice of nucleon model to study nuclear effects with this data.

In summary, this measurement of the BSA in the deeply virtual coherent exclusive 
photo-production on a nucleus is the first to clearly isolate the effect of 
coherent nuclear DVCS and of nuclear GPDs. While the statistical precision and 
the kinematic coverage are still behind the experimental results of
the proton, the results appear to match very well the predictions using the GPD 
framework. Moreover, the extraction of the CFF appears to be very convenient based
on the BSA measurement only. Together, these findings  validate the relevance of
coherent nuclear DVCS to study the nucleus globally in terms of quarks and 
gluons. 

\subsection{Incoherent DVCS}

\begin{figure}[bp!]
\center
\includegraphics[width=15cm]{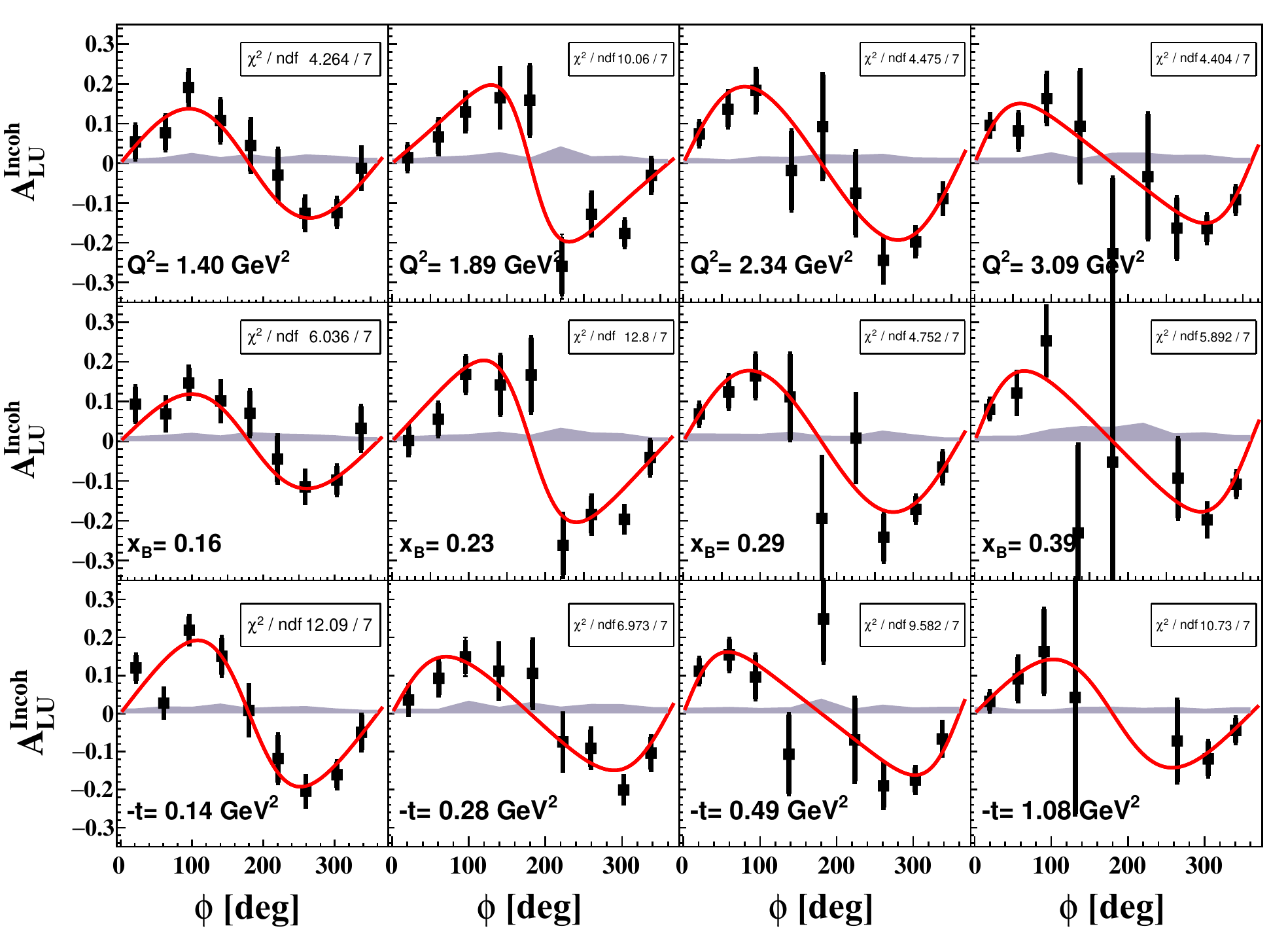}
	\caption{The BSA in the incoherent exclusive photo-production off a proton bound in
	helium-4 as a function of $\phi$ and $Q^2$ 
	(top panels), $x$ (middle panels) and $-t$ (lower panels). The error bars are  
	statistical and the gray bands represent the systematic uncertainties. The data is fitted with the 
	form $\frac{\alpha \sin(\phi)}{1+\beta \cos(\phi)}$; the results of the 
	fits are drawn with black full lines.}
\label{fig:InCohALUphi}
\end{figure}

The results for the measurement of the BSA in the incoherent DVCS channel are presented in
Fig. \ref{fig:InCohALUphi}. They display patterns rather similar to those observed with the 
free proton, with a clear domination of their sinusoidal component. To compare the data to 
models, we extract the BSA at 90 degrees with a fit of the form $\frac{\alpha \sin(\phi)}{1+\beta \cos(\phi)}$. 

\begin{figure}[tbp!]
\center
\includegraphics[width=7.4cm]{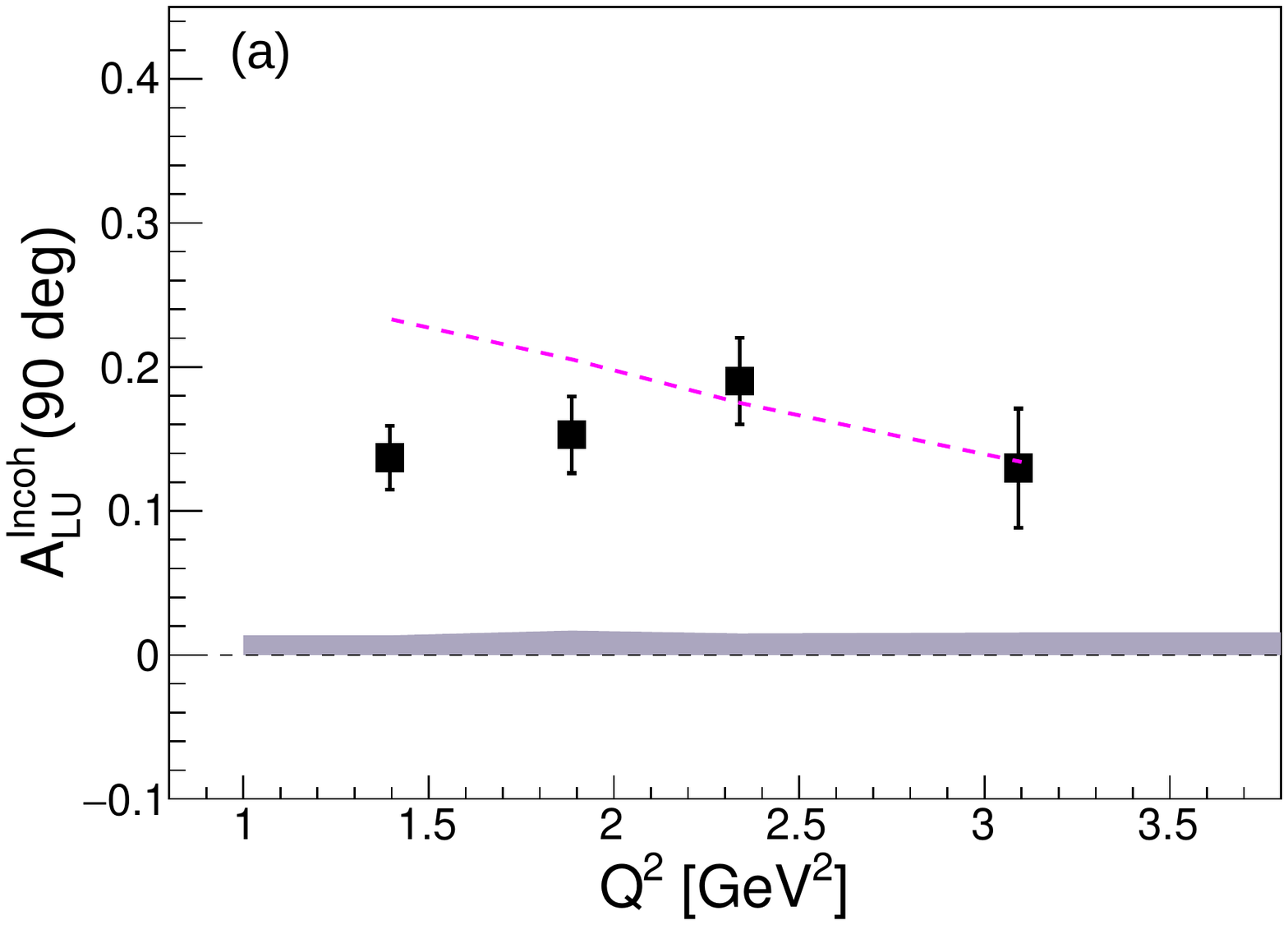}
\includegraphics[width=7.4cm]{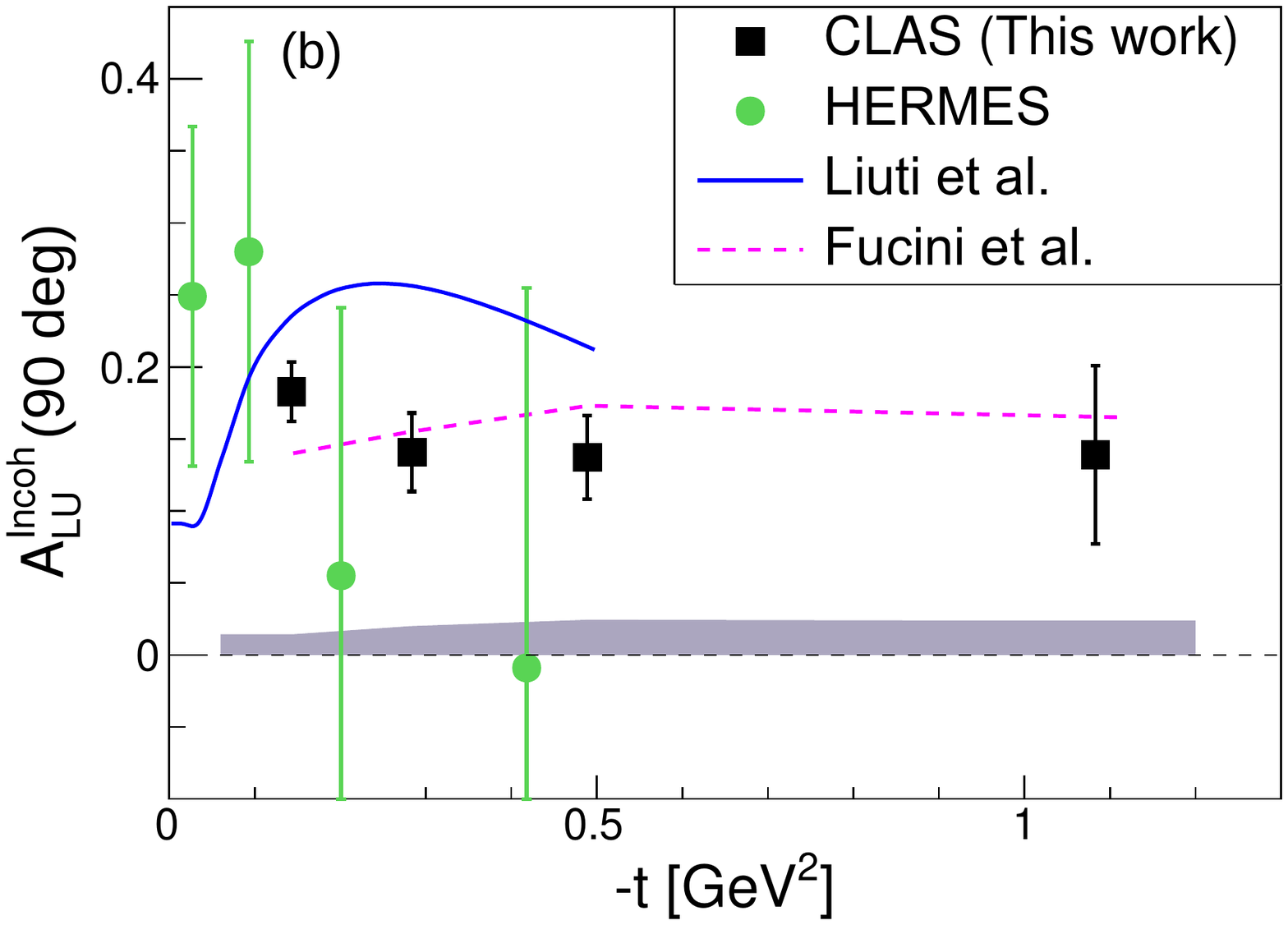}
\includegraphics[width=7.4cm]{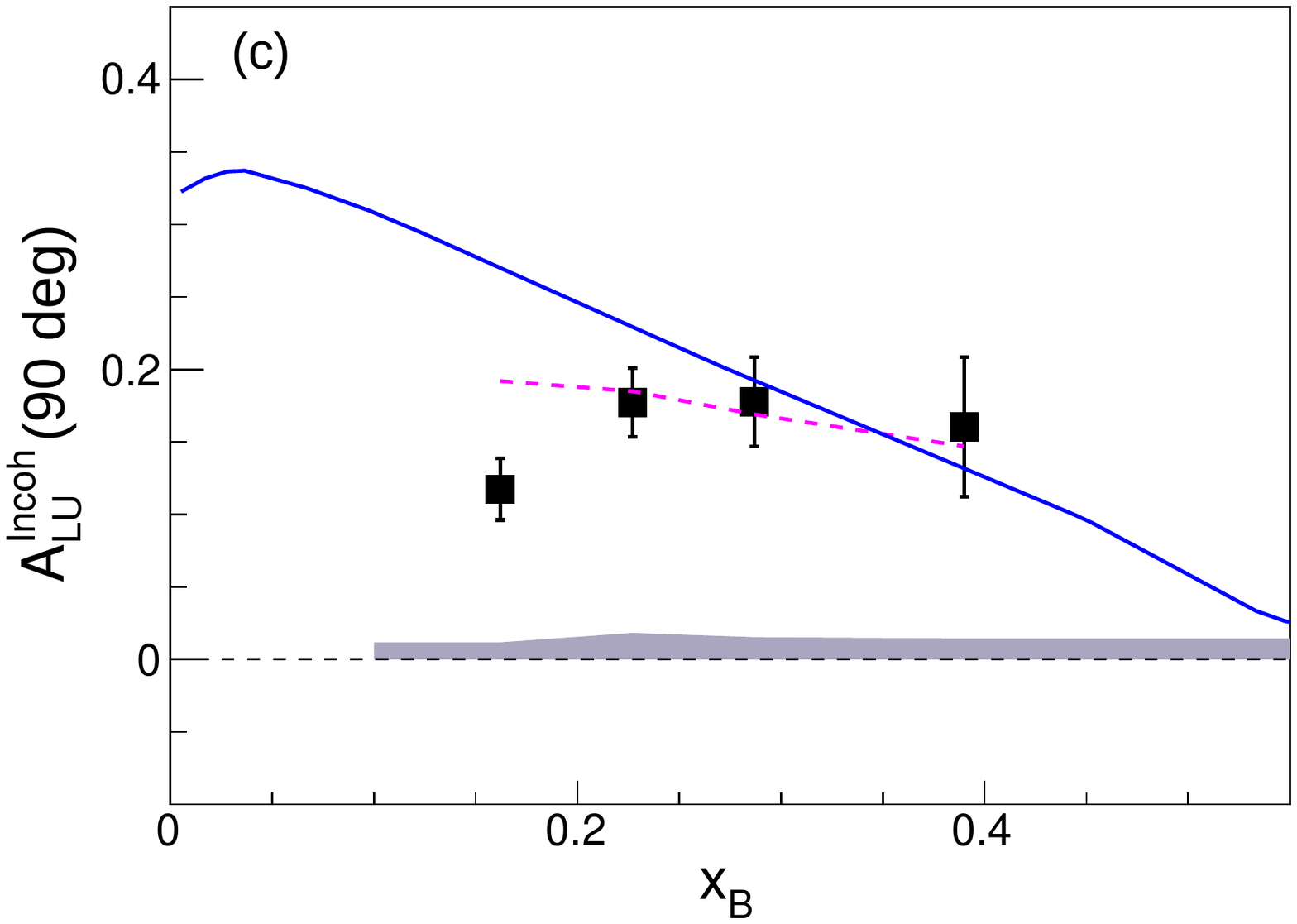}
	\caption{The BSA at 90 degrees as a function of $Q^2$ (panel a), $-t$ (panel b) 
        and $x_B$ (panel c). Our measurement is represented with black squares and the HERMES 
	measurement \cite{Airapetian:2009cga} with green circles. The theoretical prediction 
	by Liuti {\it et al.}~\cite{Liuti:2005gi,GonzalezHernandez:2012jv} is shown 
	by the full blue line, while the calculation by Fucini 
	{\it et al.}~\cite{Fucini:2019xlc} is shown with the magenta dashed line.}
\label{fig:IncALU}
\end{figure}

The asymmetries at 90 degrees are presented in Fig. \ref{fig:IncALU} together with the theoretical
calculation by the same groups as presented in Fig. \ref{fig:CohALU90}. We observe a 
significant improvement on the precision compared to 
the HERMES data, which offers more constraint on the models presented. As in the
coherent case, the calculation appears to have issues reproducing the shape of the data,
with Fucini {\it et al.}~\cite{Fucini:2019xlc} doing better than the others.
However, this time the calculations overshoot the data, sometimes by a significant amount.

\begin{figure}[tbp!]
\center
\includegraphics[width=7.4cm]{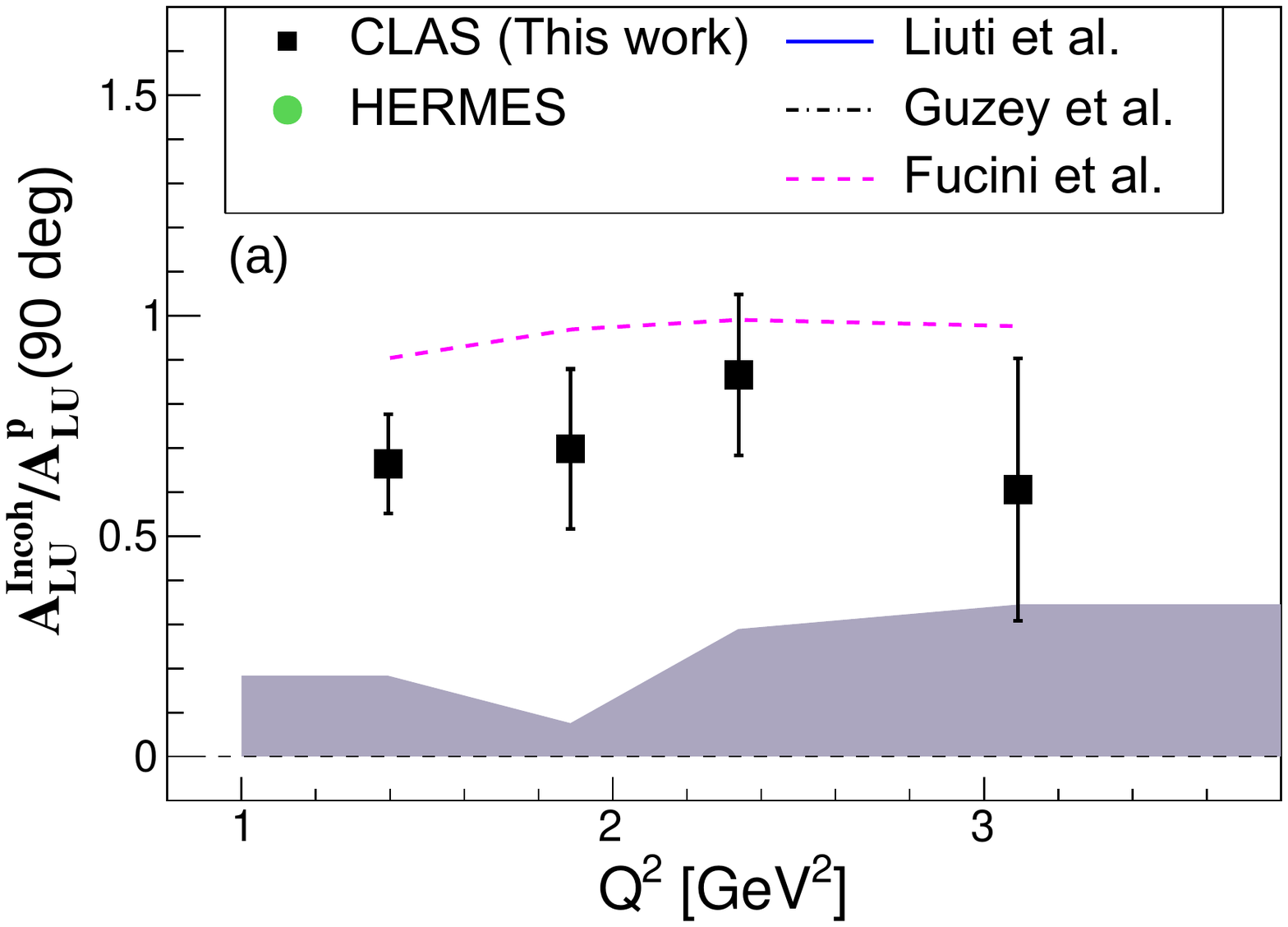}
\includegraphics[width=7.4cm]{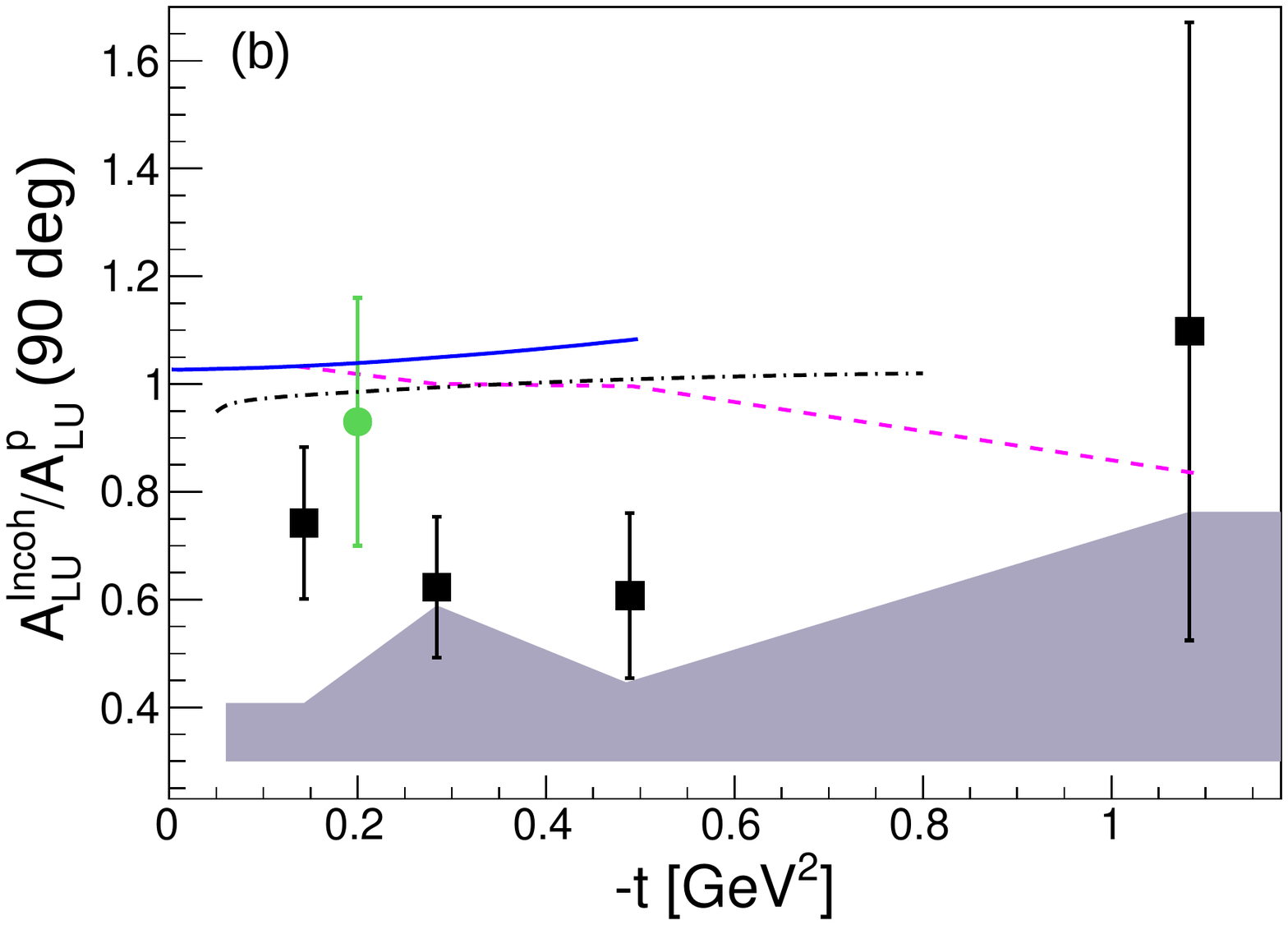}
\includegraphics[width=7.4cm]{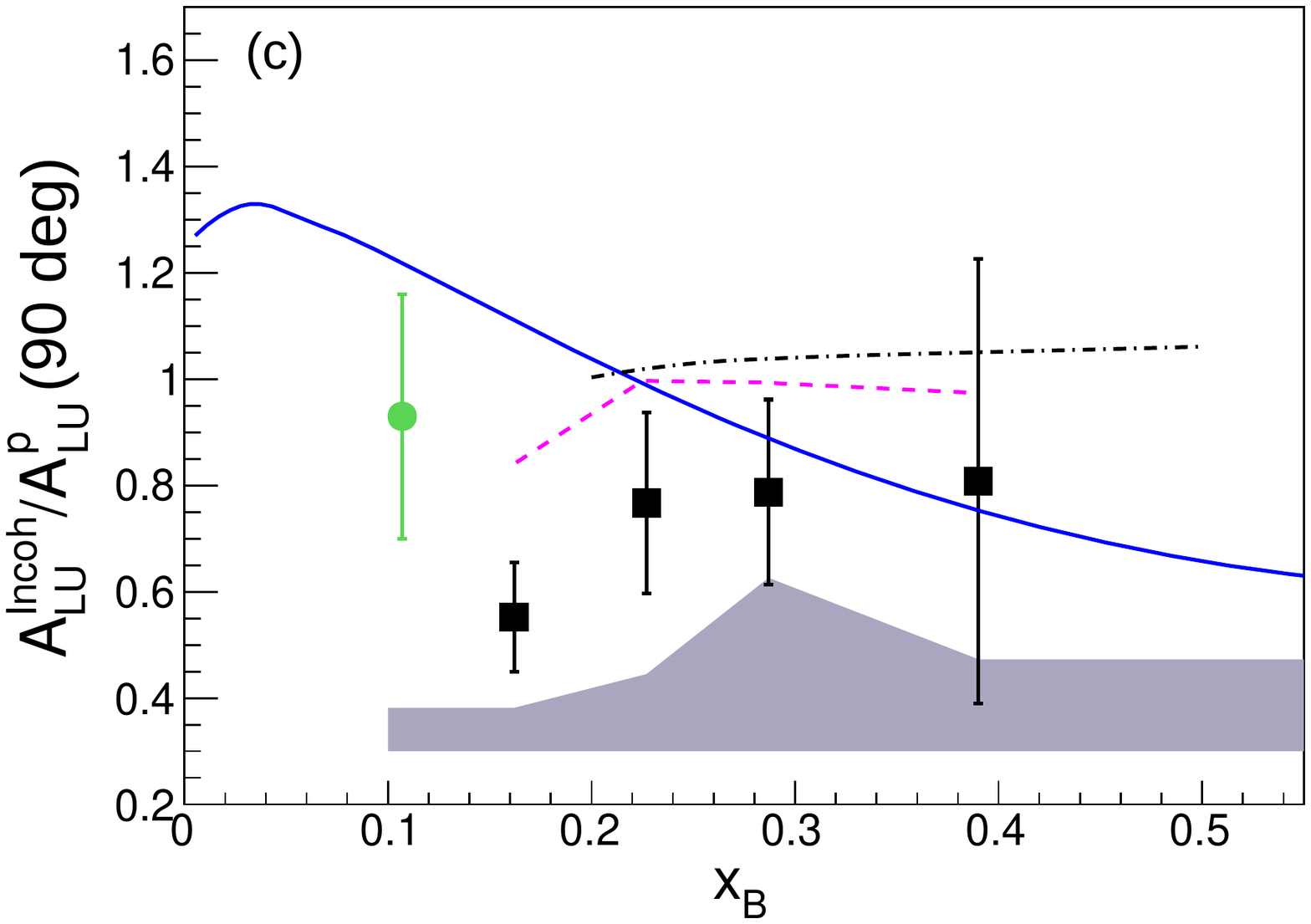}
	\caption{DVCS BSA ratio of the bound proton to the free proton as a function of 
	$Q^2$ (panel a), $-t$ (panel b) and $x_B$ (panel c). The present measurement is 
	represented with black squares and the HERMES 
	measurement \cite{Airapetian:2009cga} with green circles. The theoretical prediction
	by Liuti {\it et al.}~\cite{Liuti:2005gi,GonzalezHernandez:2012jv} is shown 
	by the full blue line, the calculation by Fucini 
	{\it et al.}~\cite{Fucini:2019xlc} is shown with the magenta dashed line, and 
	the black dot-dashed line is the calculation by Guzey {\it et al.}~\cite{Guzey:2008th}.}
\label{fig:IncRatios}
\end{figure}

An interesting way to look into this data is to show the result on incoherent nuclear DVCS compared 
with the free proton one. We can for instance make a ratio, in a fashion similar to the EMC 
effect, which allows to cancel out
the effects from the nucleon structure and highlight nuclear effects. Such a ratio is presented
in Fig. \ref{fig:IncRatios}.
Notably, the calculation by Fucini {\it et al.}~\cite{Fucini:2019xlc} appears closer than the
others with this observable. This feature indicates that the different raw asymmetry results might be 
linked to the different input model used for the free nucleon GPD rather than to differences in the 
treatment of the nuclear effects. Also, Fucini {\it et al.} appear to roughly reproduce the 
shape of the $x_B$ distribution, which might indicate that it is linked to correlations between kinematic variables.
In conclusion, the BSA in the incoherent DVCS channel is suppressed by 20 to 30\% compared to the free proton,
which was not expected by most models.

The explanation for this surprising behavior can come from different sources
both in the initial state and in the final state. 
Further work is needed to fully comprehend this newly discovered nuclear effect. On the 
experimental side, the use of tagging methods, where the nuclear fragments are measured
appear to offer the best option forward. Indeed, tagging offers the best chance to understand better
this result by offering better control over both the initial and the final state 
effects~\cite{Dupre:2015jha}.

\section{Summary}

We report the measurement of the coherent and incoherent DVCS processes off helium-4 with
CLAS at JLab. To properly isolate the coherent channel, the experiment used a specially 
designed RTPC to detect the scattered helium-4. This coherent 
DVCS measurement reveals the large BSA ($\sim 35$\%) expected by theoretical calculations
made in the impulse approximation. Moreover, we showed that the CFF extraction can be 
immediately performed using these data without any model assumptions. The incoherent DVCS measurement 
however reveals relatively small asymmetries in comparison to previous free proton 
measurements. The source of this suppression of the BSA remains unclear as both 
initial and final state effect contributions could lead to such outcome. We presented 
various models for both channels. While old and recent work agree nicely with the data
from the coherent channel, it appears more difficult to reproduce the incoherent DVCS data.
A future experimental program using tagging at the upgraded CLAS12 detector with 11~GeV
electron beam is 
planned to address this question in the coming years by using a new recoil detector 
design~\cite{Armstrong:2017zcm}.

%TODO
\section{Acknowledgments}
%\begin{acknowledgments}
The authors acknowledge the staff of the Accelerator and Physics Divisions at 
the Thomas Jefferson National Accelerator Facility who made this experiment 
possible. This work was supported in part by the Chilean Comisi\'on Nacional de 
Investigaci\'on Cient\'ifica y Tecnol\'ogica (CONICYT), by CONICYT PIA grant 
ACT1413, the Italian Instituto Nazionale di Fisica Nucleare, the French Centre 
National de la Recherche Scientifique, the French Commissariat \`a l'Energie 
Atomique, the U.S. Department of Energy under Contract No. DE-AC02-06CH11357, 
the United Kingdom Science and Technology Facilities Council (STFC), the 
Scottish Universities Physics Alliance (SUPA), the National Research Foundation 
of Korea, and the Office of Research and Economic Development at Mississippi 
State University. M.~Hattawy also acknowledges the support of the Consulat 
G\'en\'eral de France \`a J\'erusalem. This work has received funding from 
the European Research Council (ERC) under the European Union’s Horizon 2020 
research and innovation programme (Grant agreement No. 804480). The Southeastern Universities Research 
Association operates the Thomas Jefferson National Accelerator Facility for the 
United States Department of Energy under Contract No. DE-AC05-06OR23177.
%\end{acknowledgments}

\bibliographystyle{unsrt}
\bibliography{hadro.bib}

\appendix
\section{Expressions for the BSA of the Coherent DVCS}
\label{sec:eq}

We present in this appendix the detailed expressions used for Eq. \ref{eq:coh_BSA} 
and Eqs. \ref{eq:alpha1} to \ref{eq:alpha4}. These are adapted from the work of 
Kirchner and Müller \cite{Kirchner:2003wt} to match the notations and conventions used
in this work.
	
First, $\mathcal{P}_{1}(\phi)$ and $\mathcal{P}_{2}(\phi)$ are BH propagators and defined as:
\begin{align}
&\mathcal{P}_{1}(\phi) = \frac{(k - q')^{2}}{Q^{2}} = - \frac{1}{y (1 + \epsilon^{2})} 
\big[ J + 2 K \cos(\phi) \big] \\
&\mathcal{P}_{2}(\phi) = \frac{(k - \Delta)^{2}}{Q^{2}} = 1 + \frac{t}{Q^{2}} + 
\frac{1}{y (1 + \epsilon^{2})} \big[ J + 2 K \cos(\phi) \big]
\end{align}
with,
\begin{align}
& J = \bigg( 1 - y - \frac{y \epsilon^{2}}{2} \bigg) \bigg(1 + \frac{t}{Q^{2}} \bigg) - 
(1 - x_{A})(2 - y) \frac{t}{Q^{2}} \\
& K^{2} = - \delta t \, (1 - x_{A}) \bigg( 1 - y - \frac{y^{2} \epsilon^{2}}{4} \bigg) 
\bigg\{ \sqrt{1 + \epsilon^{2}} + \frac{4 x_{A} (1-x_{A}) + \epsilon^{2}}{4 (1 - x_{A})}
\delta t \bigg\} \\
& \delta t = \frac{t - t_{min}}{Q^{2}} = \frac{t}{Q^{2}} + \frac{2(1-x_{A}) \left(1- \sqrt{1 + 
\epsilon^{2}} \right) + \epsilon^{2}}{4 x_{A} (1- x_{A}) + \epsilon^{2}}.
\end{align}

The Fourier coefficients for BH contributions are defined as:
\begin{eqnarray}
c_0^{BH} = & \bigg[ & \left\{ {(2-y)}^2 + y^2{(1+\epsilon^2)}^2 \right\} 
\left\{ \frac{\epsilon^2 Q^2}{t} + 4 (1-x_A) + (4x_A+\epsilon^2) \frac{t}{Q^2} 
\right\} \nonumber \\
& \phantom{\bigg[} & + 2 \epsilon^2 \left\{ 4(1-y)(3+2\epsilon^2) + y^2(2-\epsilon^4) 
\right\} - 4 x_A^2{(2-y)}^2 (2+\epsilon^2) \frac{t}{Q^2} \nonumber \\
& \phantom{\bigg[} & + 8 K^2 \frac{\epsilon^2 Q^2}{t} \,\,\,\,\,\,\, \bigg] F_A^2(t)  \\
c_1^{BH} = & \phantom{\bigg[} & -8 (2-y) K \left\{ 2 x_A + \epsilon^2 - 
\frac{\epsilon^2 Q^2}{t} \right\} F_A^2(t)  \\
c_2^{BH} = & \phantom{\bigg[} & 8 K^2 \frac{\epsilon^2 Q^2}{t} F_A^2(t),
\end{eqnarray} 
where $F_A(t)$ is the electromagnetic form factor of $^4$He. 
The coefficient for the DVCS contribution is given by: 
\begin{equation}
   c_0^{DVCS}= 2 \frac{2-2y+y^2 + \frac{\epsilon^2}{2}y^2}{1 + \epsilon^2} \, 
   {\mathcal H}_A {\mathcal H}^{\star}_A .
   \label{eq:c0DVCS}
\end{equation}

Finally, the interference amplitude coefficients are written as:
\begin{equation}
s_{1}^{INT} = F_{A}(t) \Im m(\mathcal{H}_{A}) S_{++}(1),
\end{equation}
with
\begin{eqnarray}
   S_{++}(1) &=& \frac{-8K(2-y)y}{1+\epsilon^2} \left( 1 + 
\frac{1-x_A+\frac{\sqrt{1+\epsilon^2}-1}{2}}{1+\epsilon^2} 
\frac{t-t_{min}}{Q^{2}} \right) \label{eq:s1I}
\end{eqnarray}

\begin{eqnarray}
c_0^{INT} &=& F_A(t) \Re e(\mathcal{H}_{A}) C_{++}(0),
\end{eqnarray}
with \begin{eqnarray}  C_{++}(0) &=&
\frac{-4(2-y)(1+\sqrt{1+\epsilon^{2}})}{(1+\epsilon^{2})^2}  \bigg\{ 
   \frac{\widetilde{K}^2}{Q^2}  \frac{(2-y)^2}{\sqrt{1+\epsilon^{2}}} \, \\
   &+& \frac{t}{Q^2}  \left( 1 - y - \frac{\epsilon^2}{4} y^2 \right)  
(2-x_{A}) \left(  1 + \frac{2x_A(2-x_A + \frac{\sqrt{1+\epsilon^{2}}-1}{2} + 
\frac{\epsilon^{2}}{2x_A})\frac{t}{Q^2} + \epsilon^{2}}{(2-x_A) 
(1+\sqrt{1+\epsilon^{2}})}  \right)  \bigg\} \nonumber
 \label{eq:c0I} 
 \end{eqnarray}

\begin{eqnarray}
   c_1^{INT} &=&  F_A(t) \Re e(\mathcal{H}_{A}) C_{++}(1),
\end{eqnarray}
with  
   \begin{eqnarray}
   C_{++}(1) &=&
   \frac{-16K(1-y+\frac{\epsilon^{2}}{4}y^2)}{(1+\epsilon^{2})^{5/2}}\bigg\{\left(1+(1-x_A)\frac{\sqrt{1+\epsilon^{2}}-1}{2x_A} 
   + \frac{\epsilon^{2}}{4x_A}\right) 
\frac{x_At}{Q^2}-\frac{3\epsilon^{2}}{4.0} \bigg\} \nonumber \\&-& 4K \left( 
2-2y+y^2+\frac{\epsilon^{2}}{2}y^2\right)\frac{1+\sqrt{1+\epsilon^{2}}-\epsilon^{2}}{(1+e2)^{5/2}}
\bigg\{1-(1-3x_A)\frac{t}{Q^2}\nonumber\\&\,\,\,\,&\,\,\,\,\,\,\,\,\,\,\,\,\,\,\,\,\,\,\,\,\,+
\frac{1-\sqrt{1+\epsilon^{2}}+3\epsilon^{2}}{1+\sqrt{1+\epsilon^{2}}-\epsilon^{2}} 
\frac{x_A t}{Q^2}\bigg\}. \label{eq:c1I}
\end{eqnarray}

\section{Tables of Results with Kinematics Information}
\label{sec:fullresults}

\begin{table}[!h]
   \begin{center}
      \begin{tabular}{|c|c|c|c|c|}
         \hline
 $\langle Q^{2} \rangle$ & $\langle x_{B} \rangle$ & $\langle -t \rangle$ & $\langle \phi \rangle$ & $A_{LU}$ $\pm$ stat. $\pm$ syst.\\
 (GeV$^{2}$) &           & (GeV$^{2}$) & (degree) &  \\
         \hline
        &       &       & 24    &  0.133  $\pm$ 0.109  $\pm$ 0.026  \\ 
        &       &       & 61    &  0.321  $\pm$ 0.093  $\pm$ 0.019  \\ 
        &       &       & 99    &  0.371  $\pm$ 0.103  $\pm$ 0.040  \\ 
        &       &       & 141   &  0.245  $\pm$ 0.152  $\pm$ 0.027  \\ 
  1.14  & 0.136 & 0.096 & 178   &  0.023  $\pm$ 0.163  $\pm$ 0.028  \\ 
        &       &       & 219   & -0.053  $\pm$ 0.148  $\pm$ 0.025  \\ 
        &       &       & 263   & -0.264  $\pm$ 0.120  $\pm$ 0.039  \\ 
        &       &       & 302   & -0.176  $\pm$ 0.097  $\pm$ 0.026  \\ 
        &       &       & 338   & -0.279  $\pm$ 0.105  $\pm$ 0.034  \\ 
  \hline 
        &       &       &  21   &  0.192  $\pm$ 0.089  $\pm$ 0.027  \\ 
        &       &       &  57   &  0.282  $\pm$ 0.087  $\pm$ 0.025  \\ 
        &       &       &  97   &  0.486  $\pm$ 0.129  $\pm$ 0.043  \\ 
        &       &       & 140   &  0.100  $\pm$ 0.168  $\pm$ 0.025  \\ 
  1.42  & 0.172 & 0.099 & 180   &  0.146  $\pm$ 0.191  $\pm$ 0.030  \\ 
        &       &       & 219   & -0.111  $\pm$ 0.185  $\pm$ 0.034  \\ 
        &       &       & 263   & -0.352  $\pm$ 0.137  $\pm$ 0.037  \\ 
        &       &       & 302   & -0.414  $\pm$ 0.084  $\pm$ 0.038  \\ 
        &       &       & 338   & -0.279  $\pm$ 0.084  $\pm$ 0.026  \\ 
  \hline 
        &       &       &  21.4 &  0.180  $\pm$ 0.081  $\pm$ 0.023  \\ 
        &       &       &  57.2 &  0.350  $\pm$ 0.082  $\pm$ 0.019  \\ 
        &       &       &  96.2 &  0.270  $\pm$ 0.123  $\pm$ 0.017  \\ 
        &       &       & 139.5 &  0.305  $\pm$ 0.239  $\pm$ 0.017  \\ 
  1.90  & 0.224 & 0.107 & 178.2 &  0.103  $\pm$ 0.267  $\pm$ 0.013  \\ 
        &       &       & 221.4 & -0.212  $\pm$ 0.215  $\pm$ 0.015  \\ 
        &       &       & 263.3 & -0.306  $\pm$ 0.131  $\pm$ 0.026  \\ 
        &       &       & 303.3 & -0.138  $\pm$ 0.094  $\pm$ 0.021  \\ 
        &       &       & 338.5 & -0.163  $\pm$ 0.079  $\pm$ 0.016  \\ 
         \hline 
      \end{tabular}
      \caption{Values of the coherent $A_{LU}$ in $Q^2$ bins from Fig. \ref{fig:CohALUphi}.}
      \label{table:Coh_Q2_BSA}
   \end{center}
\end{table}                    

\begin{table}[!h]
   \begin{center}
      \begin{tabular}{|c|c|c|c|c|}
         \hline
 $\langle Q^{2} \rangle$ & $\langle x_{B} \rangle$ & $\langle -t \rangle$ & $\langle \phi \rangle$ & $A_{LU}$ $\pm$ stat. $\pm$ syst.\\
 (GeV$^{2}$) &           & (GeV$^{2}$) & (degree) &  \\
         \hline

        &       &       &  26   &  0.017  $\pm$ 0.144  $\pm$ 0.022  \\
        &       &       &  62   &  0.348  $\pm$ 0.087  $\pm$ 0.020  \\
        &       &       &  99   &  0.381  $\pm$ 0.095  $\pm$ 0.041  \\
        &       &       & 142   &  0.294  $\pm$ 0.138  $\pm$ 0.033  \\
  1.16  & 0.132 & 0.095 & 178   &  0.043  $\pm$ 0.152  $\pm$ 0.029  \\
        &       &       & 219   & -0.035  $\pm$ 0.132  $\pm$ 0.024  \\
        &       &       & 263   & -0.277  $\pm$ 0.105  $\pm$ 0.037  \\
        &       &       & 301   & -0.214  $\pm$ 0.084  $\pm$ 0.026  \\
        &       &       & 335   & -0.234  $\pm$ 0.122  $\pm$ 0.032  \\
   \hline 
        &       &       &   23  &  0.158  $\pm$ 0.085  $\pm$ 0.023  \\
        &       &       &   57  &  0.173  $\pm$ 0.088  $\pm$ 0.020  \\
        &       &       &   96  &  0.226  $\pm$ 0.133  $\pm$ 0.030  \\
        &       &       &  139  &  0.245  $\pm$ 0.176  $\pm$ 0.019  \\
  1.44  & 0.170 & 0.099 &  180  & -0.102  $\pm$ 0.192  $\pm$ 0.020  \\
        &       &       &  219  & -0.288  $\pm$ 0.191  $\pm$ 0.027  \\
        &       &       &  264  & -0.294  $\pm$ 0.136  $\pm$ 0.029  \\
        &       &       &  303  & -0.398  $\pm$ 0.092  $\pm$ 0.033  \\
        &       &       &  338  & -0.269  $\pm$ 0.083  $\pm$ 0.025  \\
   \hline 
        &       &       &  20   &  0.263  $\pm$ 0.076  $\pm$ 0.025  \\
        &       &       &  56   &  0.428  $\pm$ 0.089  $\pm$ 0.022  \\
        &       &       &  96   &  0.493  $\pm$ 0.139  $\pm$ 0.027  \\
        &       &       & 138   & -0.274  $\pm$ 0.280  $\pm$ 0.017  \\
  1.84  & 0.225 & 0.107 & 180   &  0.847  $\pm$ 0.250  $\pm$ 0.020  \\
        &       &       & 225   & -0.051  $\pm$ 0.281  $\pm$ 0.027  \\
        &       &       & 263   & -0.342  $\pm$ 0.169  $\pm$ 0.035  \\
        &       &       & 305   & -0.136  $\pm$ 0.103  $\pm$ 0.026  \\
        &       &       & 340   & -0.166  $\pm$ 0.077  $\pm$ 0.018  \\
         \hline
      \end{tabular}
      \caption{Values of the coherent $A_{LU}$ in $x_B$ bins from Fig. \ref{fig:CohALUphi}.}
      \label{table:Coh_xB_BSA}
   \end{center}
\end{table}                        

\begin{table}[!h]
   \begin{center}
      \begin{tabular}{|c|c|c|c|c|}
         \hline
 $\langle Q^{2} \rangle$ & $\langle x_{B} \rangle$ & $\langle -t \rangle$ & $\langle \phi \rangle$ & $A_{LU}$ $\pm$ stat. $\pm$ syst.\\
 (GeV$^{2}$) &           & (GeV$^{2}$) & (degree) &  \\
  \hline
        &       &       &  23   &  0.238  $\pm$ 0.093  $\pm$ 0.026  \\
        &       &       &  58   &  0.301  $\pm$ 0.087  $\pm$ 0.024  \\
        &       &       &  98   &  0.490  $\pm$ 0.112  $\pm$ 0.039  \\
        &       &       & 139   &  0.197  $\pm$ 0.160  $\pm$ 0.025  \\
  1.36  & 0.160 & 0.080 & 179   &  0.058  $\pm$ 0.192  $\pm$ 0.037  \\
        &       &       & 223   & -0.165  $\pm$ 0.164  $\pm$ 0.037  \\
        &       &       & 266   & -0.347  $\pm$ 0.134  $\pm$ 0.040  \\
        &       &       & 300   & -0.289  $\pm$ 0.093  $\pm$ 0.029  \\
        &       &       & 339   & -0.185  $\pm$ 0.086  $\pm$ 0.028  \\
  \hline 
        &       &       &  21   &  0.248  $\pm$ 0.093  $\pm$ 0.027  \\ 
        &       &       &  56   &  0.339  $\pm$ 0.083  $\pm$ 0.028  \\ 
        &       &       &  98   &  0.347  $\pm$ 0.116  $\pm$ 0.031  \\ 
        &       &       & 142   &  0.146  $\pm$ 0.189  $\pm$ 0.022  \\ 
  1.51  & 0.179 & 0.094 & 180   & -0.281  $\pm$ 0.186  $\pm$ 0.036  \\ 
        &       &       & 219   &  0.210  $\pm$ 0.200  $\pm$ 0.015  \\ 
        &       &       & 263   & -0.240  $\pm$ 0.128  $\pm$ 0.028  \\ 
        &       &       & 304   & -0.199  $\pm$ 0.096  $\pm$ 0.029  \\ 
        &       &       & 339   & -0.210  $\pm$ 0.088  $\pm$ 0.020  \\ 
  \hline
        &       &       &  22   &  0.028  $\pm$ 0.091  $\pm$ 0.021  \\
        &       &       &  61   &  0.358  $\pm$ 0.093  $\pm$ 0.020  \\
        &       &       &  97   &  0.256  $\pm$ 0.127  $\pm$ 0.031  \\
        &       &       & 140   &  0.221  $\pm$ 0.193  $\pm$ 0.020  \\
  1.61  & 0.193 & 0.127 & 179   &  0.514  $\pm$ 0.183  $\pm$ 0.035  \\
        &       &       & 218   & -0.247  $\pm$ 0.166  $\pm$ 0.019  \\
        &       &       & 261   & -0.292  $\pm$ 0.130  $\pm$ 0.033  \\
        &       &       & 303   & -0.249  $\pm$ 0.089  $\pm$ 0.028  \\
        &       &       & 337   & -0.283  $\pm$ 0.090  $\pm$ 0.026  \\
         \hline
      \end{tabular}
      \caption{Values of the coherent $A_{LU}$ in $-t$ bins from Fig. \ref{fig:CohALUphi}.}
      \label{table:Coh_t_BSA}
   \end{center}
\end{table}

% incoherent channel

\begin{table}[!h]
   \begin{center}
      \begin{tabular}{|c|c|c|c|c|}
         \hline
 $\langle Q^{2} \rangle$ & $\langle x_{B} \rangle$ & $\langle -t \rangle$ & $\langle \phi \rangle$ & $A_{LU}$ $\pm$ stat. $\pm$ syst.\\
 (GeV$^{2}$) &           & (GeV$^{2}$) & (degree) &  \\
 \hline 
        &       &        &   21   &   0.054  $\pm$  0.044   $\pm$ 0.012  \\
        &       &        &   63   &   0.077  $\pm$  0.046   $\pm$ 0.016  \\
        &       &        &   95   &   0.191  $\pm$  0.047   $\pm$ 0.026  \\
        &       &        &  140   &   0.108  $\pm$  0.056   $\pm$ 0.016  \\
  1.40  & 0.166 & 0.376  &  182   &   0.045  $\pm$  0.069   $\pm$ 0.023  \\
        &       &        &  220   &  -0.029  $\pm$  0.068   $\pm$ 0.015  \\
        &       &        &  258   &  -0.126  $\pm$  0.046   $\pm$ 0.023  \\
        &       &        &  303   &  -0.124  $\pm$  0.040   $\pm$ 0.020  \\
        &       &        &  337   &  -0.012  $\pm$  0.054   $\pm$ 0.014  \\
  \hline 
        &       &        &   20   &   0.014  $\pm$  0.036   $\pm$ 0.012  \\
        &       &        &   61   &   0.067  $\pm$  0.046   $\pm$ 0.017  \\
        &       &        &   96   &   0.130  $\pm$  0.052   $\pm$ 0.020  \\
        &       &        &  141   &   0.165  $\pm$  0.077   $\pm$ 0.029  \\
  1.89  & 0.232 & 0.415  &  180   &   0.159  $\pm$  0.089   $\pm$ 0.015  \\
        &       &        &  222   &  -0.259  $\pm$  0.081   $\pm$ 0.043  \\
        &       &        &  260   &  -0.128  $\pm$  0.056   $\pm$ 0.018  \\
        &       &        &  304   &  -0.176  $\pm$  0.039   $\pm$ 0.020  \\
        &       &        &  338   &  -0.030  $\pm$  0.045   $\pm$ 0.011  \\
  \hline 
        &       &        &   21   &   0.074  $\pm$  0.033   $\pm$ 0.014  \\
        &       &        &   58   &   0.136  $\pm$  0.046   $\pm$ 0.010  \\
        &       &        &   95   &   0.184  $\pm$  0.057   $\pm$ 0.018  \\
        &       &        &  141   &  -0.018  $\pm$  0.101   $\pm$ 0.016  \\
  2.34  & 0.288 & 0.497  &  182   &   0.092  $\pm$  0.133   $\pm$ 0.024  \\
        &       &        &  225   &  -0.075  $\pm$  0.107   $\pm$ 0.021  \\
        &       &        &  261   &  -0.244  $\pm$  0.060   $\pm$ 0.024  \\
        &       &        &  303   &  -0.198  $\pm$  0.038   $\pm$ 0.015  \\
        &       &        &  339   &  -0.089  $\pm$  0.040   $\pm$ 0.015  \\
  \hline 
        &       &        &   20   &   0.096  $\pm$  0.030   $\pm$ 0.015  \\ 
        &       &        &   57   &   0.082  $\pm$  0.048   $\pm$ 0.015  \\
        &       &        &   94   &   0.163  $\pm$  0.069   $\pm$ 0.028  \\
        &       &        &  138   &   0.093  $\pm$  0.141   $\pm$ 0.013  \\
  3.10  & 0.379 & 0.641  &  180   &  -0.227  $\pm$  0.192   $\pm$ 0.027  \\
        &       &        &  226   &  -0.033  $\pm$  0.160   $\pm$ 0.027  \\
        &       &        &  264   &  -0.163  $\pm$  0.080   $\pm$ 0.021  \\
        &       &        &  303   &  -0.164  $\pm$  0.041   $\pm$ 0.022  \\
        &       &        &  341   &  -0.091  $\pm$  0.037   $\pm$ 0.014  \\
 \hline
      \end{tabular}
      \caption{Values of the incoherent $A_{LU}$ in $Q^2$ bins from Fig. \ref{fig:InCohALUphi}.}
      \label{table:InCoh_Q2_BSA}
   \end{center}
\end{table}

\begin{table}[!h]
   \begin{center}
      \begin{tabular}{|c|c|c|c|c|}
         \hline
 $\langle Q^{2} \rangle$ & $\langle x_{B} \rangle$ & $\langle -t \rangle$ & $\langle \phi \rangle$ & $A_{LU}$ $\pm$ stat. $\pm$ syst.\\
 (GeV$^{2}$) &           & (GeV$^{2}$) & (degree) &  \\
         \hline
        &       &       &  21    &  0.094  $\pm$ 0.046   $\pm$  0.014  \\ 
        &       &       &  63    &  0.069  $\pm$ 0.044   $\pm$  0.017  \\ 
        &       &       &  96    &  0.147  $\pm$ 0.044   $\pm$  0.022  \\ 
        &       &       & 140    &  0.102  $\pm$ 0.052   $\pm$  0.015  \\ 
  1.45  & 0.163 & 0.374 & 181    &  0.071  $\pm$ 0.062   $\pm$  0.024  \\ 
        &       &       & 220    & -0.045  $\pm$ 0.062   $\pm$  0.020  \\ 
        &       &       & 259    & -0.115  $\pm$ 0.043   $\pm$  0.018  \\ 
        &       &       & 303    & -0.098  $\pm$ 0.039   $\pm$  0.015  \\ 
        &       &       & 337    &  0.033  $\pm$ 0.057   $\pm$  0.011  \\ 
  \hline
        &       &       &  22    &  0.002  $\pm$ 0.038   $\pm$  0.013  \\
        &       &       &  60    &  0.056  $\pm$ 0.044   $\pm$  0.016  \\
        &       &       &  96    &  0.168  $\pm$ 0.050   $\pm$  0.018  \\
        &       &       & 141    &  0.142  $\pm$ 0.079   $\pm$  0.025  \\
  1.93  & 0.225 & 0.381 & 182    &  0.167  $\pm$ 0.096   $\pm$  0.017  \\
        &       &       & 223    & -0.262  $\pm$ 0.083   $\pm$  0.034  \\
        &       &       & 260    & -0.185  $\pm$ 0.052   $\pm$  0.023  \\
        &       &       & 303    & -0.196  $\pm$ 0.037   $\pm$  0.021  \\
        &       &       & 337    & -0.041  $\pm$ 0.046   $\pm$  0.010  \\
  \hline
        &       &       &  21    &  0.069  $\pm$ 0.033   $\pm$  0.020  \\
        &       &       &  59    &  0.124  $\pm$ 0.046   $\pm$  0.020  \\
        &       &       &  94    &  0.165  $\pm$ 0.058   $\pm$  0.019  \\
        &       &       & 139    &  0.111  $\pm$ 0.110   $\pm$  0.024  \\
  2.33  & 0.283 & 0.468 & 181    & -0.194  $\pm$ 0.155   $\pm$  0.015  \\
        &       &       & 225    &  0.008  $\pm$ 0.111   $\pm$  0.014  \\
        &       &       & 261    & -0.242  $\pm$ 0.066   $\pm$  0.027  \\
        &       &       & 303    & -0.171  $\pm$ 0.038   $\pm$  0.018  \\
        &       &       & 338    & -0.065  $\pm$ 0.041   $\pm$  0.010  \\
  \hline
        &       &       &  20    &  0.081  $\pm$ 0.029   $\pm$  0.014  \\
        &       &       &  55    &  0.121  $\pm$ 0.054   $\pm$  0.015  \\
        &       &       &  93    &  0.253  $\pm$ 0.090   $\pm$  0.031  \\
        &       &       & 135    & -0.230  $\pm$ 0.225   $\pm$  0.039  \\
  2.98  & 0.389 & 0.688 & 180    & -0.052  $\pm$ 0.425   $\pm$  0.036  \\
        &       &       & 231    & -0.377  $\pm$ 0.334   $\pm$  0.047  \\
        &       &       & 266    & -0.093  $\pm$ 0.103   $\pm$  0.020  \\
        &       &       & 303    & -0.198  $\pm$ 0.045   $\pm$  0.023  \\
        &       &       & 341    & -0.108  $\pm$ 0.035   $\pm$  0.016  \\
 \hline
 \end{tabular}
 \caption{Values of the incoherent $A_{LU}$ in $x_B$ bins from Fig. \ref{fig:InCohALUphi}.}
 \label{table:InCoh_xB_BSA}
 \end{center}
\end{table}

\begin{table}[!h]
   \begin{center}
      \begin{tabular}{|c|c|c|c|c|}
         \hline
 $\langle Q^{2} \rangle$ & $\langle x_{B} \rangle$ & $\langle -t \rangle$ & $\langle \phi \rangle$ & $A_{LU}$ $\pm$ stat. $\pm$ syst.\\
 (GeV$^{2}$) &           & (GeV$^{2}$) & (degree) &  \\
  \hline
        &       &        &   22    &  0.120   $\pm$  0.037  $\pm$ 0.014  \\ 
        &       &        &   61    &  0.027   $\pm$  0.042  $\pm$ 0.019  \\ 
        &       &        &   96    &  0.219   $\pm$  0.041  $\pm$ 0.018  \\ 
        &       &        &  142    &  0.150   $\pm$  0.054  $\pm$ 0.026  \\ 
  1.84  & 0.215 & 0.135  &  180    &  0.008   $\pm$  0.067  $\pm$ 0.015  \\ 
        &       &        &  221    & -0.119   $\pm$  0.065  $\pm$ 0.018  \\ 
        &       &        &  260    & -0.204   $\pm$  0.043  $\pm$ 0.020  \\ 
        &       &        &  303    & -0.160   $\pm$  0.037  $\pm$ 0.014  \\ 
        &       &        &  338    & -0.049   $\pm$  0.047  $\pm$ 0.010  \\ 
  \hline 
        &       &        &   21    &  0.036   $\pm$  0.040  $\pm$ 0.013  \\
        &       &        &   61    &  0.093   $\pm$  0.045  $\pm$ 0.013  \\
        &       &        &   96    &  0.149   $\pm$  0.051  $\pm$ 0.034  \\
        &       &        &  139    &  0.111   $\pm$  0.073  $\pm$ 0.018  \\
  2.15  & 0.257 & 0.281  &  183    &  0.105   $\pm$  0.093  $\pm$ 0.030  \\
        &       &        &  223    & -0.074   $\pm$  0.076  $\pm$ 0.018  \\
        &       &        &  259    & -0.091   $\pm$  0.055  $\pm$ 0.025  \\
        &       &        &  302    & -0.200   $\pm$  0.039  $\pm$ 0.025  \\
        &       &        &  338    & -0.104   $\pm$  0.047  $\pm$ 0.017  \\
  \hline 
        &       &        &   21    &  0.111   $\pm$  0.037  $\pm$ 0.015  \\
        &       &        &   60    &  0.154   $\pm$  0.045  $\pm$ 0.018  \\
        &       &        &   94    &  0.096   $\pm$  0.060  $\pm$ 0.015  \\
        &       &        &  138    & -0.107   $\pm$  0.105  $\pm$ 0.017  \\
  2.37  & 0.291 & 0.492  &  183    &  0.248   $\pm$  0.119  $\pm$ 0.039  \\
        &       &        &  224    & -0.069   $\pm$  0.110  $\pm$ 0.013  \\
        &       &        &  261    & -0.190   $\pm$  0.062  $\pm$ 0.023  \\
        &       &        &  303    & -0.174   $\pm$  0.036  $\pm$ 0.016  \\
        &       &        &  338    & -0.067   $\pm$  0.047  $\pm$ 0.018  \\
  \hline 
        &       &        &   20    &  0.032   $\pm$  0.030  $\pm$ 0.018  \\
        &       &        &   57    &  0.091   $\pm$  0.058  $\pm$ 0.011  \\
        &       &        &   91    &  0.163   $\pm$  0.112  $\pm$ 0.011  \\
        &       &        &  131    &  0.042   $\pm$  0.307  $\pm$ 0.018  \\
  2.45  & 0.312 & 1.089  &  175    & -0.936   $\pm$  0.397  $\pm$ 0.018  \\
        &       &        &  231    & -1.189   $\pm$  0.517  $\pm$ 0.015  \\
        &       &        &  264    & -0.072   $\pm$  0.109  $\pm$ 0.017  \\
        &       &        &  305    & -0.119   $\pm$  0.048  $\pm$ 0.013  \\
        &       &        &  341    & -0.044   $\pm$  0.036  $\pm$ 0.016  \\
 \hline
 \end{tabular}
 \caption{Values of the incoherent $A_{LU}$ in $-t$ bins from Fig. \ref{fig:InCohALUphi}.}
 \label{table:InCoh_t_BSA}
 \end{center}
\end{table}

\begin{table}[!h]
   \begin{center}
      \begin{tabular}{|c|c|c|c|}
         \hline
 $\langle Q^{2} \rangle$ & $\langle x_{B} \rangle$ & $\langle -t \rangle$  & $A_{LU}$(90 deg) $\pm$ stat. $\pm$ syst.\\
 (GeV$^{2}$) &           & (GeV$^{2}$) &  \\
  \hline
  1.14  & 0.136 & 0.096 &  0.304  $\pm$ 0.051  $\pm$ 0.032 \\
  1.42  & 0.172 & 0.099 &  0.364  $\pm$ 0.059  $\pm$ 0.037 \\
  1.90  & 0.224 & 0.107 &  0.295  $\pm$ 0.061  $\pm$ 0.028 \\
  \hline 
  1.16  & 0.132 & 0.095 &  0.320  $\pm$ 0.045  $\pm$ 0.038 \\
  1.44  & 0.17  & 0.099 &  0.278  $\pm$ 0.079  $\pm$ 0.027 \\
  1.84  & 0.225 & 0.107 &  0.320  $\pm$ 0.161  $\pm$ 0.037 \\
  \hline 
  1.36  & 0.160 & 0.080 &  0.376  $\pm$ 0.042  $\pm$ 0.033 \\
  1.51  & 0.179 & 0.094 &  0.245  $\pm$ 0.072  $\pm$ 0.031 \\
  1.61  & 0.193 & 0.127 &  0.318  $\pm$ 0.095  $\pm$ 0.035 \\
  \hline
  \end{tabular}
  \caption{Values of the coherent $A_{LU}$(90 deg) in $Q^2$ (top block), $x_B$ (middle block), and $-t$ (bottom block) bins. }
  \label{table:Coh_BSA_90}
  \end{center}
\end{table}

% incoherent channel

\begin{table}[!h]
\begin{center}
\begin{tabular}{|c|c|c|c|}
\hline
 $\langle Q^{2} \rangle$ & $\langle x_{B} \rangle$ & $\langle -t \rangle$  & $A_{LU}$(90 deg) $\pm$ stat. $\pm$ syst.\\
 (GeV$^{2}$) &           & (GeV$^{2}$) &  \\
 \hline 
  1.40  & 0.166 & 0.376 & 0.137  $\pm$ 0.022  $\pm$ 0.014 \\   
  1.89  & 0.232 & 0.415 & 0.153  $\pm$ 0.027  $\pm$ 0.017 \\   
  2.34  & 0.288 & 0.497 & 0.190  $\pm$ 0.030  $\pm$ 0.017 \\   
  3.10  & 0.379 & 0.641 & 0.130  $\pm$ 0.041  $\pm$ 0.016 \\   
 \hline 
  1.45  & 0.163 & 0.374 & 0.117  $\pm$ 0.021  $\pm$ 0.012 \\   
  1.93  & 0.225 & 0.381 & 0.177  $\pm$ 0.024  $\pm$ 0.018 \\   
  2.33  & 0.283 & 0.468 & 0.178  $\pm$ 0.031  $\pm$ 0.015 \\   
  2.98  & 0.389 & 0.688 & 0.160  $\pm$ 0.048  $\pm$ 0.014 \\   
 \hline 
  1.84  & 0.215 & 0.135 & 0.183  $\pm$ 0.021  $\pm$ 0.014 \\   
  2.15  & 0.257 & 0.281 & 0.141  $\pm$ 0.027  $\pm$ 0.020 \\   
  2.37  & 0.291 & 0.492 & 0.137  $\pm$ 0.029  $\pm$ 0.024 \\   
  2.45  & 0.312 & 1.089 & 0.139  $\pm$ 0.062  $\pm$ 0.024 \\   
 \hline
 \end{tabular}
 \caption{Values of the incoherent $A_{LU}$(90 deg) in $Q^2$ (top block), $x_B$ (middle block), and $-t$ (bottom block) bins.}
 \label{table:InCoh_BSA_90}
 \end{center}
\end{table}

\end{document}